\documentclass[lettersize,journal]{IEEEtran}
\usepackage{amsmath,amsfonts}
\usepackage{algorithmic}
\usepackage{algorithm}
\usepackage{array}
\usepackage[caption=false,font=normalsize,labelfont=sf,textfont=sf]{subfig}
\usepackage{textcomp}
\usepackage{stfloats}
\usepackage{url}
\usepackage{verbatim}
\usepackage{graphicx}
\usepackage{cite}

\usepackage{multirow}
\usepackage{amsmath}
\usepackage{amssymb}
\usepackage{booktabs}
\usepackage{makecell}
\usepackage[mathscr]{eucal}
\DeclareSubrefFormat{myparens}{#1(#2)}
\begin{document}

\title{Uplink Assisted Joint Channel Estimation and CSI Feedback: An Approach Based on Deep Joint Source-Channel Coding}

\author{Yiran Guo, Wei Chen,~\IEEEmembership{Senior Member,~IEEE,} Bo Ai,~\IEEEmembership{Fellow,~IEEE}

\thanks{Yiran Guo, Wei Chen and Bo Ai are with School of Electronic and Information Engineering, Beijing Jiaotong University, Beijing, China. (e-mail: \{yiranguo, weich, boai\}@bjtu.edu.cn)}
}


\maketitle

\begin{abstract}
% In the frequency division duplex (FDD) multiple-input multiple-output (MIMO) wireless communication systems, the acquisition of downlink channel state information (CSI) is essential for maximizing spatial resource utilization and improving system spectral efficiency. This paper simultaneously addresses the tasks of channel estimation (CE) and CSI feedback in downlink CSI acquisition, proposing a deep learning-based joint CE and CSI feedback network. It takes into account challenges in the traditional source-channel separate coding (SCCC) architecture, as well as the partial reciprocity between the uplink and downlink channels in FDD systems. The proposed network is integrated with the deep joint source-channel coding (DJSCC) architecture, incorporating uplink CSI as auxiliary information for end-to-end network training, which enhances CSI accuracy. Experimental results demonstrate that the proposed network overcomes the ``cliff effect'' in the separate architecture, benefiting from the DJSCC architecture to extract more effective semantic information and achieve higher CSI reconstruction accuracy. Furthermore, the impact of non-ideal uplink channel estimation on system performance is discussed, along with a network training strategy to mitigate its effects. The effectiveness of uplink CSI as assisted information and the necessity of an end-to-end multi-module joint training architecture is validated through comprehensive ablation and scalability experiments.

In frequency division duplex (FDD) multiple-input multiple-output (MIMO) wireless communication systems, the acquisition of downlink channel state information (CSI) is essential for maximizing spatial resource utilization and improving system spectral efficiency. The separate design of modules in AI-based CSI feedback architectures under traditional modular communication frameworks, including channel estimation (CE), CSI compression and feedback, leads to sub-optimal performance. In this paper, we propose an uplink assisted joint CE and and CSI feedback approach via deep learning for downlink CSI acquisition, which mitigates performance degradation caused by distribution bias across separately trained modules in traditional modular communication frameworks. The proposed network adopts a deep joint source-channel coding (DJSCC) architecture to mitigate the ``cliff effect'' encountered in the conventional separate source-channel coding. Furthermore, we exploit the uplink CSI as auxiliary information to enhance CSI reconstruction accuracy by leveraging the partial reciprocity between the uplink and downlink channels in FDD systems, without introducing additional overhead. The effectiveness of uplink CSI as assisted information and the necessity of an end-to-end multi-module joint training architecture is validated through comprehensive ablation and scalability experiments.

%Experimental results demonstrate that the proposed network overcomes the ``cliff effect'' in the separate architecture, benefiting from the end-to-end DJSCC-based architecture, and achieves higher CSI reconstruction accuracy. Furthermore, the impact of non-ideal uplink channel estimation on system performance is discussed, along with a network training strategy. 
\end{abstract}

\begin{IEEEkeywords}
CSI feedback, channel estimation, frequency division duplex, deep joint source-channel coding.
\end{IEEEkeywords}

\section{Introduction}
The successful deployment of fifth-generation (5G) wireless communication systems across multiple domains has spurred interest in next-generation wireless technologies. With the growing demand for extended reality—encompassing virtual reality, mixed reality, and augmented reality \cite{XR}—alongside applications such as telemedicine, smart cities, and autonomous driving, sixth-generation (6G) wireless networks must support higher communication capacity and increasingly diverse services. To meet the more stringent key performance indicators (KPIs) of 6G, ultra-massive multiple-input multiple-output (MIMO) technology, which has already demonstrated significant spectral efficiency gains in 5G, is regarded as a promising solution due to its ability to efficiently utilize spatial resources \cite{promising_5G}.

To fully exploit the performance gains of MIMO technology, the increasing number of antennas must be complemented by accurate channel state information (CSI) acquisition at the base station (BS). In time-division duplex (TDD) systems, where the uplink and downlink channels share the same frequency, downlink CSI can be readily inferred from the estimated uplink channel due to channel reciprocity. In contrast, frequency-division duplex (FDD) systems face significant challenges in CSI acquisition, as channel reciprocity is weakened by the large frequency gap between the uplink and downlink channels. Consequently, additional spectral resources are often required to feed the estimated downlink CSI from the user equipment (UE) to the BS. Although some studies attempt to infer downlink CSI from the uplink channel \cite{UL_Pre_DL, UL_P_Rre_DL}, practical performance limitations or additional information requirements have led existing research and standardization efforts to prioritize performance gains by allocating spectral resources for CSI feedback \cite{CSI_overview, AI_CSI_usecase}. Simultaneously, efforts continue to minimize feedback overhead in CSI feedback schemes to achieve an optimal trade-off between performance and overhead. However, as the number of antennas increases, conventional CSI feedback techniques—such as compressed sensing (CS) methods based on channel sparsity assumptions and codebook-based approaches used in 5G—struggle to achieve satisfactory CSI reconstruction performance under limited feedback resource constraints. The integration of artificial intelligence (AI) offers a promising solution to this challenge. AI-based CSI compression techniques leverage CSI correlations across spatial, temporal, and frequency domains using an autoencoder architecture, enabling efficient CSI compression, low-overhead feedback, and accurate reconstruction\cite{CSI_overview}. 

In \cite{CSINET}, a CSI feedback architecture based on convolutional neural networks (CNNs) was proposed. Building on this, a network architecture that accounts for quantization errors, called CsiNet+, was introduced in \cite{CsiNet_Plus}. Similarly, an enhanced Swin Transformer-based architecture was employed in \cite{Swim_Transformer} to improve CSI reconstruction accuracy. These approaches treat CSI as the source information and replace the traditional source compression module in communication systems with an autoencoder, and the conventional separate channel coding module is applied. This framework is referred to as the separate source-channel coding (SSCC) architecture. However, this architecture suffers from the ``cliff effect'', i.e., the performance experiences a significant drop when the channel condition deteriorates. The ``cliff effect'' can be avoided by using deep joint source-channel coding (DJSCC) architecture \cite{ADJSCC}. The DJSCC encoder extracts low-dimensional semantic representations of CSI, which are then fed back to reduce overhead while mitigating channel fading and ensuring high-quality CSI reconstruction at the receiver via the DJSCC decoder. The DJSCC-based CSI feedback architecture has been extended to multi-user joint CSI feedback and precoding in \cite{JFPNet}. Additionally, the signal-to-noise ratio (SNR) information is novelly introduced as auxiliary information to enable the DJSCC network adaptable to varying channel conditions \cite{ADJSCC}. Leveraging auxiliary information to improve reconstruction accuracy has been considered in several studies \cite{Bi-Directional, CA_Net, DNN_TypeII, Multi_Domain_Assisted}. In \cite{Bi-Directional}, partial reciprocity between uplink and downlink CSI is exploited in FDD systems by considering scattered environments and fixed channel parameters, allowing uplink CSI to assist downlink CSI reconstruction. However, many existing works overlook the impact of increasing antenna numbers on downlink channel estimation (CE) and the corresponding limitations of CE resources. This joint CE and CSI feedback architecture reduces channel estimation overhead and simultaneously addresses the distribution mismatch problem commonly found in traditional AI-based modular communication systems. In such systems, individual modules are often designed and trained independently and later integrated directly for testing. However, the output distribution of one module may differ from the training distribution of the subsequent module, leading to degraded system performance. Studies in \cite{CA_Net, ICC_JEFPNet, YuWei_JEFPNet} jointly consider downlink CE and CSI feedback within SSCC architectures to mitigate the impact of CE errors. Additionally, in \cite{JSCC_CE_FB}, CE and CSI feedback are jointly designed under the DJSCC architecture, utilizing the second-order statistical properties of the channel to optimize power allocation. Nevertheless, this approach ignores the potential performance gains from the partial reciprocity between uplink and downlink CSI in DJSCC architecture and assumes ideal uplink CE under an additive white Gaussian noise (AWGN) channel without fading.

In this paper, we propose an uplink-assisted joint design of downlink CE and CSI feedback within the DJSCC framework. Our approach integrates downlink pilot design, CE, and CSI feedback while considering the partial reciprocity characteristics of the channel in FDD systems. We introduce uplink CSI as auxiliary information for CSI reconstruction, training the network in an end-to-end manner to optimize the parameters of different modules. This ensures that only the key information differing from the uplink CSI is compressed and fed back at the UE, maximizing key information retention for a fixed feedback overhead. Meanwhile, the estimated uplink CSI at the BS is utilized to compensate for missing information of the reciprocal component. In summary, the main contributions of this paper are summarized as follows:

\begin{itemize}
\item[$\bullet$]We propose a joint CE and CSI feedback architecture under the DJSCC framework, which integrates downlink channel estimation, CSI compression and feedback. This architecture enables low-overhead CE and CSI feedback while mitigating the "cliff effect" inherent in traditional SSCC architectures and alleviating the distribution bias issue associated with multi-module separation design.

\item[$\bullet$]We evaluate the impact of uplink CE errors on CSI feedback performance under the DJSCC architecture through simulation experiments. We evaluate the impact of uplink channel estimation (CE) under various network training and testing strategies and propose a non-ideal CE and CSI feedback training strategy that accounts for uplink CE errors to mitigate the resulting performance degradation. This strategy enhances the robustness of CSI reconstruction accuracy by reducing the influence of estimation errors on the optimization objective.

\item[$\bullet$]Leveraging the partial reciprocity between uplink and downlink channels in FDD systems, we introduce an uplink-assisted joint CE and CSI feedback network. To further enhance downlink CSI reconstruction accuracy, we design a feature-merging module, referred to as the ``\emph{Joint Refine}'' module, which effectively captures and utilizes CSI reciprocity between uplink and downlink CSI.
\end{itemize}

Section \ref{section1} presents the system model of the FDD MIMO-OFDM system considered in this paper and describes the two primary communication processes: CE and CSI feedback. Section \ref{section2} provides an overview of existing network architectures for CE estimation and CSI feedback. It then introduces a joint CE and CSI feedback network based on a joint source-channel coding architecture, leveraging the partial reciprocity of uplink and downlink channels. A detailed description of the network parameters and architecture is also provided. In Section \ref{section3}, we evaluate the performance of the proposed network through simulations using the generated dataset. Finally, Section \ref{section4} summarizes our work.

\textit{Notations: }In this paper, vectors and matrices are denoted using boldface lowercase and uppercase letters, respectively. The complex conjugate transpose and transpose of a matrix $\mathbf{A}$ are represented as $\mathbf{A}^\mathrm{H}$ and $\mathbf{A}^\mathrm{T}$, respectively. The Frobenius norm and the modulus of the complex matrix are denoted by ${\left\| \cdot \right\|_\mathrm{F}}$, and $\left| \cdot \right|$, respectively. The statistical expectation is represented by $\mathbb{E}(\cdot)$, while the upward rounding operation is denoted by $\lceil{A}\rceil$. A set containing multiple elements is denoted by $\{\cdot, \cdot\}$. A set of $K$ elements from $\mathbf{A}_1$ to $\mathbf{A}_K$ is represented as $\{\mathbf{A}_k\}_{k=1}^K$, while $\{\mathbf{A}(k)\}_{k=1}^K$ indicates a set of $K$ elements from $\mathbf{A}(1)$ to $\mathbf{A}(K)$. In Section \ref{section1}, the OFDM symbol index in the temporal domain is represented in parentheses, while the subcarrier index in the frequency domain is given as a subscript. For example, $\mathbf{A}_n(m)$ denotes the value of the matrix $\mathbf{A}$ at the $n$-th subcarrier and the $m$-th OFDM symbol.  

% $\odot$ is hadamard product. $diag\left({\cdot}\right)$ denotes the diagonal matrix formed by the numbers in parentheses, and the numbers in parentheses are the diagonal elements.

\section{System model}
\label{section1}
In this paper, we consider a MIMO-OFDM system with $M$ subcarriers, where the BS is equipped with $N_{\mathrm{BS}}$ antennas and the UE has $N_{\mathrm{UE}}$ antennas. In this paper, we focus on the CE and CSI feedback components of the wireless communication system. Specifically, the CE includes the estimation of downlink CSI at the UE side, which serves as the source information for feedback, and the uplink CE, which is used as the received feedback vector.
\subsection{Channel Estimation}
Pilot-assisted methods are employed for both downlink CE and uplink CE. We employ $L$ consecutive OFDM symbols to estimate the channel matrix set $\mathscr{H} = \left\{\mathbf{H}(m)\right\}_{m=1}^M$, where $\mathbf{H}(m) \in \mathbb{C}^{A_\mathrm{r} \times A_{\mathrm{t}}}$ represents the CSI matrix for the $m$-th subcarrier. Here, $A_{\mathrm{r}}$ and $A_{\mathrm{t}}$ denote the number of antennas at the receiver and transmitter, respectively. In downlink CE, $A_{\mathrm{r}} = N_{\mathrm{UE}}$ and $A_{\mathrm{t}} = N_{\mathrm{BS}}$, while in uplink CE, $A_{\mathrm{r}} = N_{\mathrm{BS}}$ and $A_{\mathrm{t}} = N_{\mathrm{UE}}$. We assume that the channel remains time-invariant within these $L$ consecutive OFDM symbols. Each OFDM symbol consists of $M$ subcarriers, with $N_p$ subcarriers allocated for pilot transmission at pilot intervals $g$, leaving $M - N_p$ subcarriers for data or control information transmission. At the $m$-th subcarrier, the signal transmitted from the transmitter to the receiver in the spatial-temporal domain is denoted as $\mathbf{X}(m) \in \mathbb{C}^{A_{\mathrm{t}} \times L}$. The received signal at the $m$-th subcarrier, denoted as $\mathbf{Y}(m) \in \mathbb{C}^{A_{\mathrm{r}} \times L}$, can be expressed as:
\begin{equation}
    \label{Recieve_signal}
    \mathbf{Y}(m)=\mathbf{H}(m)\mathbf{X}(m)+\mathbf{N}(m),
\end{equation}
where $\mathbf{N}(m) \in \mathbb{C}^{A_{\mathrm{r}} \times L}$ represents the AWGN with mean zero and variance $\sigma^2$ per element. To distinguish between the uplink and downlink channels, we denote their respective noise variances as $\sigma_{\mathrm{u}}^2$ and $\sigma_{\mathrm{d}}^2$. For simplicity, we assume that these $L$ OFDM symbols are exclusively used for CE, meaning the data symbol position is set to 0. The transmitted signals at the $l$-th symbol are represented as:
\begin{equation}
    \label{pilot}
    \mathbf{X}_l=\mathbf{T}\mathbf{P}_l,
\end{equation}
where $\mathbf{P}_l \in \mathbb{C}^{N_p \times A_{\mathrm{t}}}$ is the pilot matrix for the $l$-th OFDM symbol, and $N_p = \lceil \frac{M}{g} \rceil$ represents the number of subcarriers allocated to the pilot with pilot intervals $g$. Each row of $\mathbf{P}_l$ satisfies the power constraints:
\begin{equation}
    \label{pilot_Power_Constrain}
    \Vert \mathbf{P}_l{(n_p)} \Vert_2^2=1 \ \left(n_p \in \left\{{1,\cdots, N_p}\right\}\right).
\end{equation} 
$\mathbf{T} \in \mathbb{R}^{M \times N_p}$ is the transformation matrix consisting of binary elements, i.e., 0 or 1, used to achieve equal interval placement of pilot symbols across all subcarriers. The matrix $\mathbf{X}_l = \left[\mathbf{x}_{l}(1), \cdots, \mathbf{x}_{l}(M)\right]^{\mathrm{T}} \in \mathbb{C}^{M \times A_{\mathrm{t}}}$ represents the transmitted signals in the spatial-frequency domain, where $\mathbf{x}_{l}(m)$ is the column vector corresponding to the $l$-th OFDM symbol of the $m$-th subcarrier, extracted from the $l$-th column of $\mathbf{X}(m)$.

The channel can be estimated using any CE algorithm $f_{\mathrm{CE}}\left(\cdot\right)$, which takes the received signals and the known pilot symbols as inputs. The estimated channel can be expressed as:
\begin{equation}
    \label{CE}
    \tilde {\mathscr{H}}=f_{CE}\left({\mathscr{Y}, \mathscr{P}}\right),
\end{equation}
where $\mathscr{Y} = \left\{\mathbf{Y}(m)\right\}_{m=1}^M$ denote the set of received symbols in the spatial-temporal-frequency domain, consisting of received signal matrix $\mathbf{Y}(m)$ in the temporal-frequency domain. Let $\mathscr{P} = \left\{\mathbf{P}_l\right\}_{l=1}^L$ represent the set of pilot symbols in the same domain, consisting of transmitted pilot matrix $\mathbf{P}_l$ in the spatial-frequency domain. The pilot sets for the uplink and downlink channels are denoted as $\mathscr{P}_{\mathrm{u}} = \{\mathbf{P}_{\mathrm{u},l}\}_{l=1}^L$ and $\mathscr{P}_{\mathrm{d}} = \{\mathbf{P}_{\mathrm{d},l}\}_{l=1}^L$, respectively. The estimated CSI matrix set is represented by $\tilde{\mathscr{H}} = \{\tilde{\mathbf{H}}(m)\}_{m=1}^{M}$. With pilot-assisted CE, the estimated uplink and downlink channels can be expressed as $\tilde{\mathscr{H}}_{\mathrm{u}} = \{\tilde{\mathbf{H}}_{\mathrm{u}}(m)\}_{m=1}^M$ and $\tilde{\mathscr{H}}_{\mathrm{d}} = \{\tilde{\mathbf{H}}_{\mathrm{d}}(m)\}_{m=1}^M$, respectively.

\subsection{CSI Compression and Feedback}
\label{Section_CSI_FB}
After the downlink CE, the estimated downlink CSI set $\tilde{\mathscr{H}}_{\mathrm{d}}$ can be treated as the source information. CSI compression is performed through the encoder module, as follows:
\begin{equation}
	\label{encoder}
\mathbf{S}=\mathcal{E}_{\alpha}\left({\tilde {\mathscr{H}}_{\mathrm{d}}}\right),
\end{equation}
where $\mathcal{E}_{\alpha}\left(\cdot\right)$ is the encoder with a trainable parameter set $\alpha$, used to compress a high-dimensional matrix into a low-dimensional complex matrix ${\mathbf{S}} = \left[\mathbf{s}(1), \cdots, \mathbf{s}(K)\right]^{\mathrm{T}} \in \mathbb{C}^{K \times N_{\mathrm{UE}}}$, satisfying $KN_{\mathrm{UE}} \ll M N_{\mathrm{BS}}N_{\mathrm{UE}}$. Here, $K$ denotes the number of subcarriers assigned for CSI feedback during the uplink transmission, while the remaining subcarriers can be used for transmitting other data, which is not considered in this paper. $\mathbf{s}(k) \in \mathbb{C}^{N_{\mathrm{UE}}}$ represents the transmit signal at the $k$-th subcarrier, corresponding to the UE's $N_{\mathrm{UE}}$ transmit antennas.

After power normalization, as $\Vert \mathbf{s}(k) \Vert_2^2=1$, where $k \in \{1,\cdots, L\}$, and OFDM mapping, compressed CSI is transmitted from the UE to the BS. The received signal after passing through the uplink feedback channel at the $k$-th uplink subcarrier can be represented as
\begin{equation}
    \label{CSI_FB_receive}
    {\mathbf{y}}(k)=\mathbf{H}_{\mathrm{u}}(k)\mathbf{s}(k)+\mathbf{n}_{\mathrm{u}}(k),
\end{equation}
where $\mathbf{n}_{\mathrm{u}}(k) \in \mathbb{C}^{N_{\mathrm{BS}}}$ represents the AWGN with mean zero and variance $\sigma_{\mathrm{u}}^2$, similar to $\mathbf{N}(m)$ in Eq.~\eqref{Recieve_signal}. In this paper, we assume that interference among various subcarriers is negligible, as it is not the focus of this research. $\mathbf{H}_{\mathrm{u}}(k) \in \mathbb{C}^{N_{\mathrm{BS}} \times N_{\mathrm{UE}}}$ denotes the uplink CSI of the $k$-th subcarrier.

% During the CSI feedback and CE, we assume that the channel is a time-invariant channel, which can be estimated at the BS through a pilot-assisted CE method:
% \begin{equation}
% 	\label{channel_est}
% {\mathbf{H}}_{\mathrm{u}} = f_{CE}\left({\mathbf{x}_p, \mathbf{Y}_p}\right),
% \end{equation}
% which $\mathbf{x}_p$ is the transmitted pilot signal with the fixed pilot interval $v$. When employing a comber pilot design, the number of transmitted pilot symbols can be denoted as $N_p$. $\mathbf{Y}_p \in \mathbb{C}^{N_p \times N_{\mathrm{BS}}}$ is the receiver pilot signal at BS. 

Using the estimated uplink CSI $\tilde{\mathscr{H}}_{\mathrm{u}}$, the transmitted symbols can be detected by the signal detection algorithm $\mathcal{F}\left(\cdot\right)$, as follows:
\begin{equation}
    \label{detection_algorithm}
    \hat{\mathbf{S}} = \mathcal{F}\left({\mathbf{S}, \tilde{\mathscr{H}}_{\mathrm{u}}}\right),
\end{equation}
which $\hat{\mathbf{S}} \in \mathbb{C}^{K \times N_{\mathrm{UE}}}$ is the received symbols corresponding to transmitted symbols $\mathbf{S}$ after uplink channel.
% \begin{equation}
%     \label{receiver}
%     {\hat{s}}_i = \mathbf{w}_i^\mathrm{H} \mathbf{y}_i,
% \end{equation}
% where $\mathbf{w}_i$ is the combiner vector by MRC:
% \begin{equation}
% 	\label{MRC}
% \mathbf{w}_i = \frac{{\hat{\mathbf{h}}_{\mathrm{u}}^{\mathrm{T}}}(i)}{\left\| {{\hat{\mathbf{h}}_{\mathrm{u}}^{\mathrm{T}}}(i)} \right\|_2} \in \mathbb{C}^{N_{\mathrm{BS}}},
% \end{equation}
% where $\hat{\mathbf{h}}(i)$ is the $i$-th row of $\hat{\mathbf{H}}_{\mathrm{u}}$, which means the uplink channel for the $i$-th subcarrier.

After the signal detection, the downlink CSI can be reconstructed using the decoder, as follows:
\begin{equation}
	\label{decoder}
\hat {\mathscr{H}}_{\mathrm{d}}=\mathcal{D}_{\beta}\left({\hat {\mathbf{S}}}\right),
\end{equation}
where $\mathcal{D}_{\beta}(\cdot)$ is the decoder function with parameter set $\beta$. $\hat {\mathscr{H}}_{\mathrm{d}} = \{{\hat{\mathbf{H}}_{\mathrm{d}}(m)}\}_{m=1}^{M}$ is reconstructed downlink CSI set with $M$ subcarriers.

\section{Uplink-Aided Joint CE and CSI Feedback Network}
\label{section2}

\begin{figure}[!t]
  \centering
  \subfloat[]{
    \label{SEFNet}
    \includegraphics[width=3.3in]{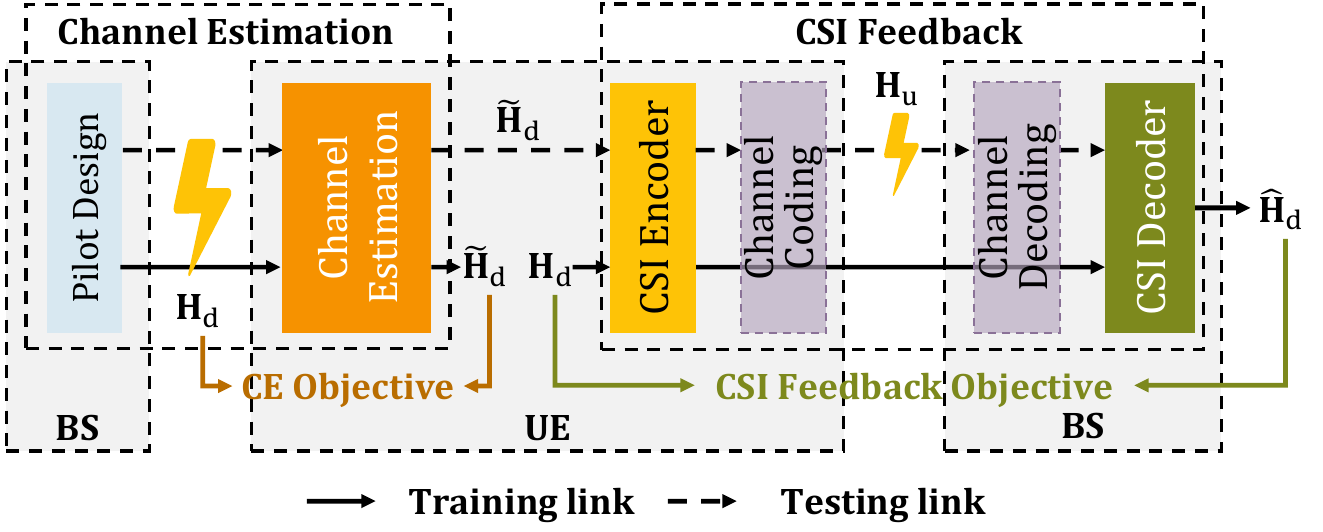}
    
  }\\
  \subfloat[]{
    \label{JEFNet_SSCC}
    \includegraphics[width=3.3in]{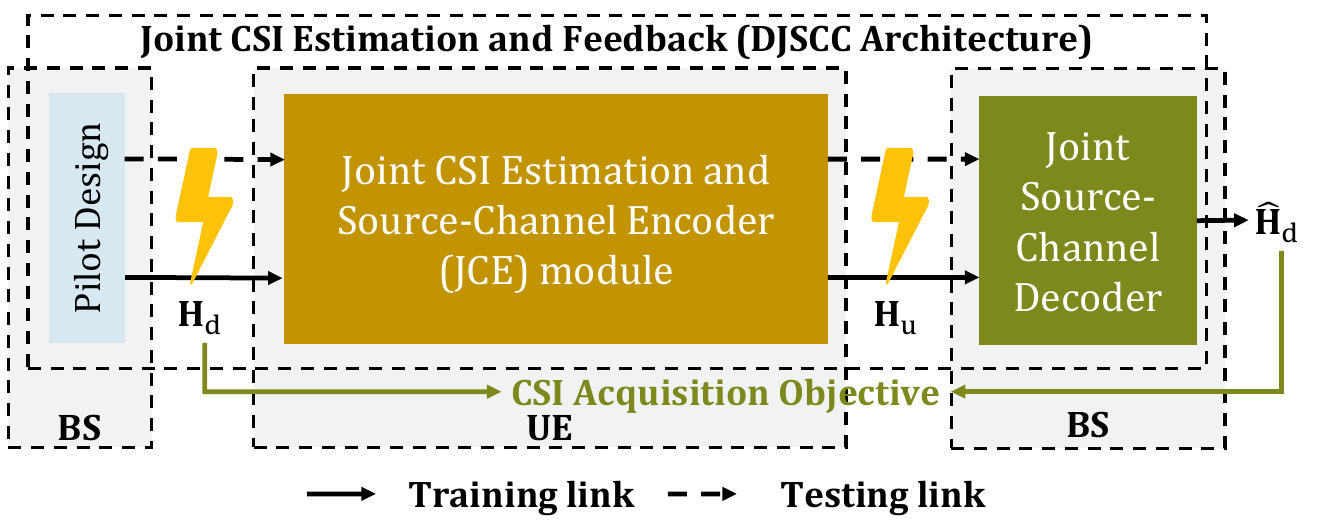}
    
  }\\
  \subfloat[]{
    \label{JEFNet_JSCC}
    \includegraphics[width=3.3in]{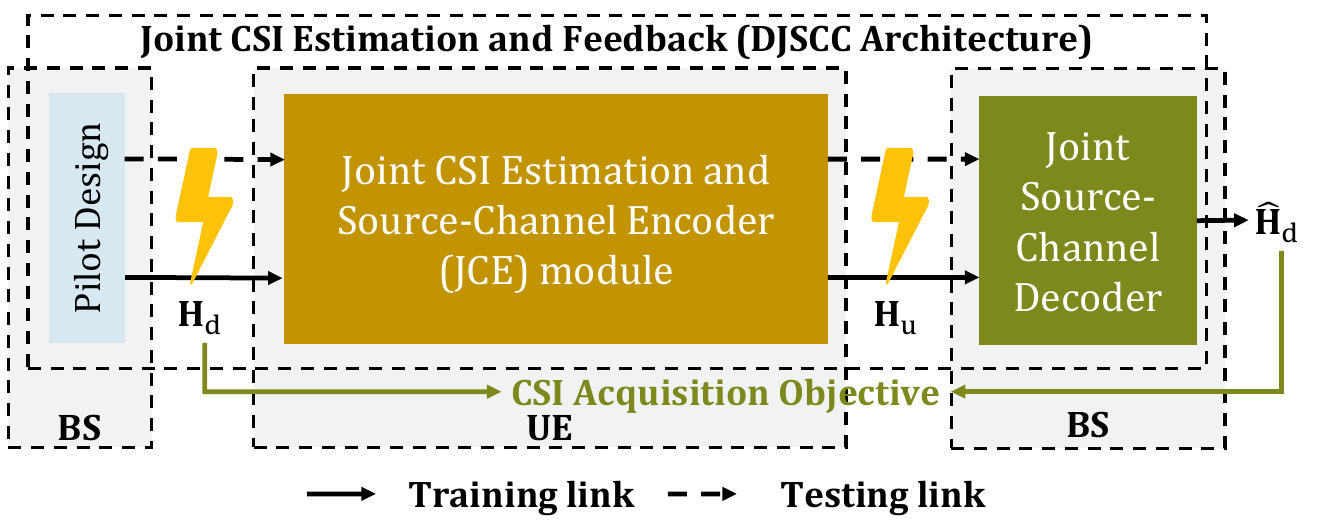}
    
  }\\
  \caption{The construction of AI-based CE and CSI feedback: \protect\subref{SEFNet} separate CE and CSI feedback; \protect\subref{JEFNet_SSCC} joint CE and CSI feedback with SSCC; \protect\subref{JEFNet_JSCC} joint CE and CSI feedback with DJSCC.}
  \label{construction}
\end{figure}

\begin{figure*}[!t]
	\centering
	\includegraphics[width=6.8in]{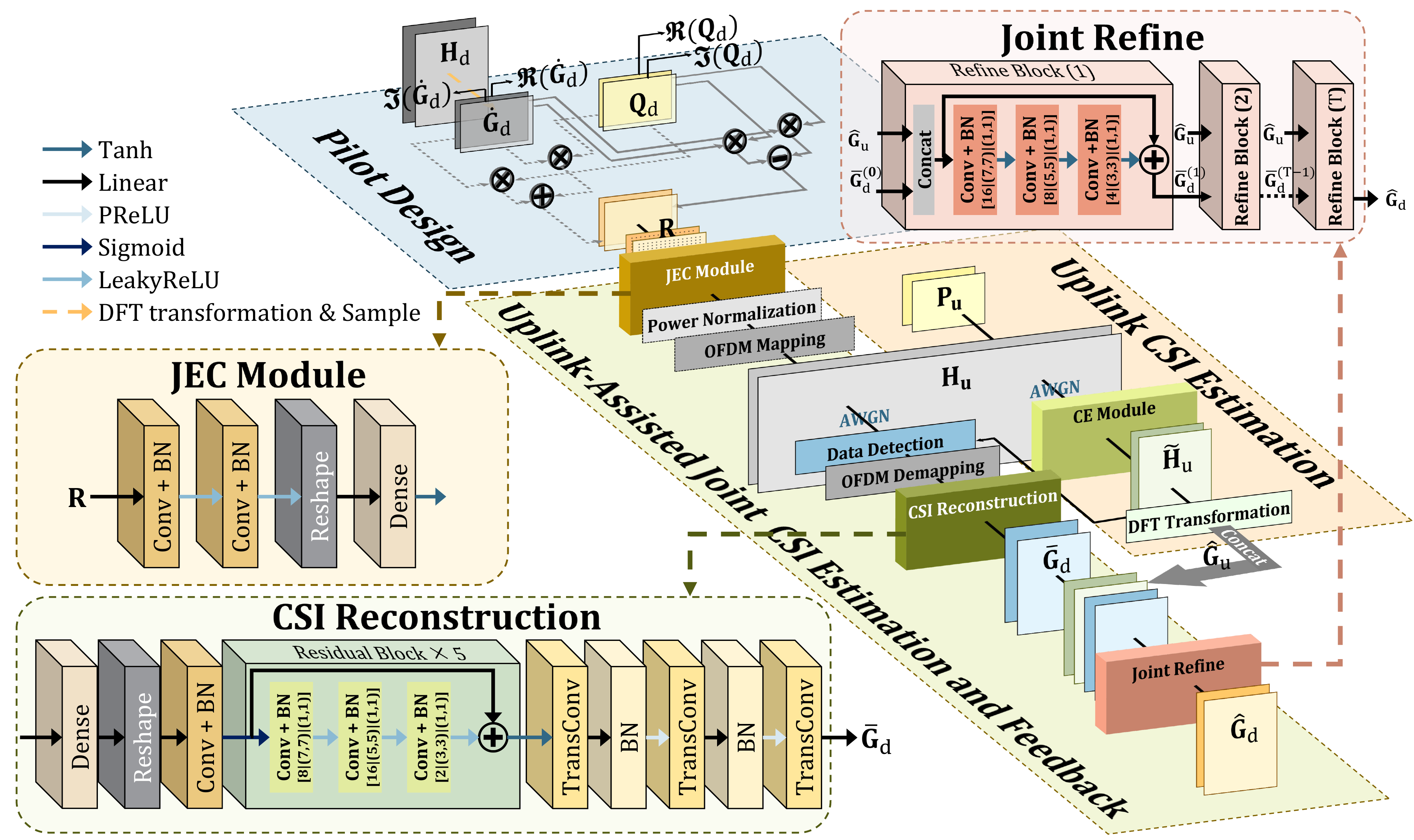}
	\caption{The network architecture of UJEFNet is illustrated with detailed figures of key modules. For clarity, some operations that do not involve trainable parameters are omitted from the figure, and parameter settings for certain network layers are summarized in Table~\ref{net_config}. For convolutional and transposed convolutional layers, respectively denoted as ``$\mathrm{Conv}$'' and ``$\mathrm{TransConv}$'', we use the notation ``$F|(W_1, W_2)|(S_1, S_2)$'', where $F$ represents the number of filters, $W_1 \times W_2$ denotes the window size, and $S_1 \times S_2$ indicates the step size. For fully connected layers, denoted as ``$\mathrm{Dense}$'', Table~\ref{net_config} specifies the number of neurons. }
	\label{UJEFNet}
\end{figure*}
Traditional communication systems employ a modular architecture, where each module is designed independently and cascaded. This architecture is also commonly used in studies of CSI feedback tasks, where the CSI feedback module and CE are separate modules. In the separate CE and CSI feedback (SEFNet) architecture, AI-based CE and CSI feedback modules are trained to enhance downlink CE accuracy and downlink CSI reconstruction accuracy, respectively, as depicted by ``$\text{CE Objective}$'' and ``$\text{CSI Feedback Objective}$'' in Fig.~\subref*{SEFNet}. Nonetheless, the isolated training approach prevents subsequent modules from understanding the distribution of preceding modules' outputs—distributions, which may differ with the training datasets in the SEFNet. Furthermore, the optimal solutions obtained for individual modules under separate objectives may not align with the overall end-to-end optimization goal. The disparity between the dataset distribution and the optimization objective will negatively impact the system's end-to-end performance.

To address the challenges posed by modular architectures, some studies propose a joint design for CE and CSI compression at the UE side, resulting in a joint CE and CSI compression module, as shown in Fig.~\subref*{JEFNet_SSCC}. In this approach, CSI features are directly extracted from the received downlink pilot signal and compressed for feedback, and the entire network is trained end-to-end to reconstruct the downlink CSI, with ``$\text{CSI Acquisition Objective}$'' as the optimization goal at the BS. We refer to this approach, which omits the effect of channel coding during training but applies conventional channel coding during testing, as the SSCC architecture. However, the SSCC architecture may suffer from the ``cliff effect'' when deployed in practice. To this end, we consider a joint CE and CSI feedback network (JEFNet) under the DJSCC architecture, as shown in Fig.~\subref*{JEFNet_JSCC}. In this architecture, the joint CE and source-channel coding (JEC) module at the UE directly extracts feature information of the downlink CSI from the pilot signal and transmits it to the joint source-channel decoder at the BS. The decoder then reconstructs the downlink CSI from the received feature information. The network is optimized using ``$\text{CSI Acquisition Objective}$'', with both uplink and downlink transmissions trained end-to-end to enable the DJSCC architecture to learn channel fading characteristics and mitigate the ``cliff effect''.

In addition, leveraging the partial reciprocity between uplink and downlink channels, we utilize uplink CSI to assist in the reconstruction of downlink CSI at the BS. This approach reduces feedback overhead and improves reconstruction accuracy. We refer to this novel network as the uplink-assisted joint CE and CSI feedback network (UJEFNet), with its architecture illustrated in Fig.~\ref{UJEFNet}. The UJEFNet comprises several modules, including pilot design, CE, source coding and decoding, and channel coding and decoding. It leverages uplink CSI to assist in downlink CSI reconstruction, with the entire system trained in an end-to-end manner. For simplicity in the description of the design, we assume that the UE has a single antenna, i.e., $N_{\mathrm{UE}} = 1$. Then, the matrix sets should be presented as vector sets, i.e., $\tilde{\mathscr{H}}_{\mathrm{u}}=\{{\tilde{\mathbf{h}}_{\mathrm{u}}(m)}\}_m^M$, $\tilde{\mathscr{H}}_{\mathrm{d}}=\{{\tilde{\mathbf{h}}_{\mathrm{d}}(m)}\}_m^M$ and $\hat{\mathscr{H}}_{\mathrm{d}}=\{{\hat{\mathbf{h}}_{\mathrm{d}}(m)}\}_m^M$. If $N_{\mathrm{UE}} > 1$, the network architecture can be extended using methods such as parallel processing or 3-dimensional convolution. 

\subsection{Joint CE and CSI Feedback with DJSCC}
In pilot-assisted channel estimation for MIMO systems, orthogonal pilots suffer from high spectral resource overhead due to the large number of antennas, whereas non-orthogonal pilots incur significant computational complexity and latency owing to the complexity of the associated channel estimation algorithms. Therefore, the design of effective pilot signals and corresponding low-complexity CE algorithms is crucial for accurate CE with minimal pilot resource overhead. 

In the SEFNet architecture for MIMO systems, in addition to facing the problems of high pilot resource overhead and high computational complexity of the CE algorithm, inaccurate CE exacerbates the mismatch between the output distribution of the CE module and the training dataset (ideal CSI) used in training the CSI feedback module, leading to performance degradation. Conversely, a more complex CE module increases the computational complexity for the UE. Therefore, jointly designing pilot signals, low-complexity CE and CSI feedback, is essential for limited pilot resource overhead. An AI-based end-to-end architecture can jointly optimize parameters across multiple modules, and reduce computational complexity at the UE. Here, we proposed a joint CE and source-channel coding module to complete CE and CSI compression, which is represented as the JEC module in Fig.~\ref{UJEFNet}.

We use a pilot design method in the angular-frequency domain based on downlink CSI to enable joint CE and CSI feedback, leveraging partial reciprocity between uplink and downlink channels. This partial reciprocity will be interpreted in the following subsection. Given that only a subset of subcarriers is used for CE, we first sample from the downlink CSI set $\mathscr{H}_{\mathrm{d}}=\left\{{\mathbf{h}_{\mathrm{d}}(m)}\right\}_{m=1}^{M}$, where $\mathbf{h}_{\mathrm{d}}(m) \in \mathbb{C}^{1 \times N_{\mathrm{BS}}}$ represents the downlink CSI matrix for the $m$-th subcarrier.
% Sequentially, only the subcarriers containing the pilot signal are remained to form a new downlink CSI matrix $\dot{\mathbf{H}}_{\mathrm{d}} \in \mathbb{C}^{N_p \times N_{\mathrm{BS}}}$ in spatial-frequency domain, which is used as the input to the pilot design network. 
Sequentially, only the subcarriers containing the pilot signals are retained to form a new downlink CSI matrix, which can be expressed as $\mathbf{T}^{\mathrm{T}} \mathbf{H}_{\mathrm{d}} \in \mathbb{C}^{N_p \times N_t}$, where ${\mathbf{H}}_{\mathrm{d}}=\left[{\mathbf{h}_{\mathrm{d}}^{\mathrm{T}}(1), \cdots, \mathbf{h}_{\mathrm{d}}^{\mathrm{T}}(M)}\right]^{\mathrm{T}} \in \mathbb{C}^{M \times N_{\mathrm{BS}}}$ denotes the spatial-frequency domain downlink CSI matrix consisting of $M$ vectors from the set $\mathscr{H}_{\mathrm{d}}$. Using the discrete Fourier transform (DFT) matrix $\mathbf{F} \in \mathbb{C}^{N_{\mathrm{BS}} \times N_{\mathrm{BS}}}$, the downlink CSI in the angular-frequency domain can be expressed as:
\begin{equation}
    \label{DFT_trans_H_DL}
     \dot{{\mathbf{G}}}_{\mathrm{d}}=\mathbf{T}^{\mathrm{T}}{\mathbf{H}}_{\mathrm{d}}\mathbf{F}.
\end{equation}
With the definition $\dot{{\mathbf{H}}}_{\mathrm{d}}=\mathbf{T}^{\mathrm{T}}{\mathbf{H}}_{\mathrm{d}} \in \mathbb{C}^{N_p \times N_t}$, it will satisfy:
\begin{equation}
    \label{DFT_trans_h_DL}
     \dot{{\mathbf{h}}}_{\mathrm{d}}(n_p)=\dot{{\mathbf{g}}}_{\mathrm{d}}(n_p)\mathbf{F}^{\mathrm{H}},
\end{equation}
where $\dot{{\mathbf{g}}}_{\mathrm{d}}(n_p)$ and $\dot{{\mathbf{h}}}_{\mathrm{d}}(n_p)$ denote the $n_p$-th row of $\dot{{\mathbf{G}}}_{\mathrm{d}}$ and $\dot{{\mathbf{H}}}_{\mathrm{d}}$, respectively.

The pilot signal received at the $n_p$-th subcarrier, where $n_p \in \{1, \cdots, N_p\}$, can be expressed as: 
\begin{equation}
\label{Recieve_AI_pilot}
   \begin{split}
       \mathbf{r}(n_p)&=\dot{\mathbf{h}}_{\mathrm{d}}(n_p)\mathbf{P}_{\mathrm{d}}(n_p)^{\mathrm{T}}+\mathbf{n}_{\mathrm{d}}(n_p)\\&=\dot{\mathbf{g}}_{\mathrm{d}}(n_p) \mathbf{F}^{\mathrm{H}} \mathbf{P}_{\mathrm{d}}(n_p)^{\mathrm{T}}+\mathbf{n}_{\mathrm{d}}(n_p)\\&
       \triangleq \dot{{\mathbf{g}}}_{\mathrm{d}}(n)\mathbf{Q}_{\mathrm{d}}(n_p)^{\mathrm{T}}+\mathbf{n}_{\mathrm{d}}(n_p),
   \end{split} 
\end{equation}
where the matrix $\mathbf{P}_{\mathrm{d}}(n_p) \in \mathbb{C}^{L \times N_{\mathrm{BS}}}$ represents the downlink pilot matrix in the spatial-temporal domain, while $\mathbf{Q}_{\mathrm{d}}(n_p) \in \mathbb{C}^{L \times N_{\mathrm{BS}}}$ corresponds to the pilot in the angular-temporal domain, both of them correspond to the $n_p$ subcarriers. The $l$-th row of $\mathbf{P}_{\mathrm{d}}(n_p)$ corresponds to the $n_p$-th row of $\mathbf{P}_{\mathrm{d}, l}$ in the set $\mathscr{P}_{\mathrm{d}}$. Additionally, $\dot{\mathbf{g}}_{\mathrm{d}} \in \mathbb{C}^{1 \times N_{\mathrm{BS}}}$ denotes the $n$-th row of $\dot{\mathbf{G}}_{\mathrm{d}}(n_p)$, and $\mathbf{n}_{\mathrm{d}}(n_p) \in \mathbb{C}^{1 \times L}$ is the additive white Gaussian noise (AWGN) with mean zero and variance $\sigma_{\mathrm{d}}^2$. $\mathbf{r}(n_p)$ denote received downlink pilots at the $n_p$-th subcarrier, while the whole received downlink pilots can be expressed as $\mathbf{R}=\left[{\mathbf{r}(1)^{\mathrm{T}},\cdots,\mathbf{r}(N_p)^{\mathrm{T}}}\right]^{\mathrm{T}} \in \mathbb{C}^{N_p \times L}$.

We define the downlink pilot set $\mathscr{Q}_{\mathrm{d}} = \{\mathbf{Q}_{\mathrm{d}}(n_p)\}_{n_p=1}^{N_p}$ in the angular-temporal-frequency domain as a set of trainable parameters with power constraints given as Eq.~\eqref{pilot_Power_Constrain}. The network is trained end-to-end to optimize the pilot parameters for a specific task objective. Since the updated parameters must be real numbers, the real and imaginary components of each complex number are treated as separate real values for individual updates. The real and imaginary parts of the received signal can be expressed as follows:
\begin{equation}
\label{Real_Recieve_AI_pilot}
\begin{split}
    \mathfrak{R}\left({\mathbf{r}(n_p)}\right)=&\mathfrak{R}\left({{\mathbf{n}_{\mathrm{d}}}(n_p)}\right) + \mathfrak{R}\left({\dot{{\mathbf{g}}}_{\mathrm{d}}(n_p)}\right)\mathfrak{R}\left({{\mathbf{Q}}_{\mathrm{d}}(n_p)^{\mathrm{T}}}\right)\\&-\mathfrak{I}\left({\dot{{\mathbf{g}}}_{\mathrm{d}}(n_p)}\right)\mathfrak{I}\left({{\mathbf{Q}}_{\mathrm{d}}(n_p)^{\mathrm{T}}}\right),
\end{split}
\end{equation}
\begin{equation}
\label{Imag_Recieve_AI_pilot}
\begin{split}
    \mathfrak{I}\left({\mathbf{r}(n_p)}\right)=&\mathfrak{I}\left({{\mathbf{n}_{\mathrm{d}}}(n_p)}\right) + \mathfrak{R}\left({\dot{{\mathbf{g}}}_{\mathrm{d}}(n_p)}\right)\mathfrak{I}\left({{\mathbf{Q}}_{\mathrm{d}}(n_p)^{\mathrm{T}}}\right)\\&+\mathfrak{I}\left({\dot{{\mathbf{g}}}_{\mathrm{d}}(n_p)}\right)\mathfrak{R}\left({{\mathbf{Q}}_{\mathrm{d}}(n_p)^{\mathrm{T}}}\right),
\end{split}
\end{equation}
where $\mathfrak{R}(\cdot)$ and $\mathfrak{I}(\cdot)$ denote the real and imaginary parts of a complex number, respectively.

Upon receiving the pilot signal, the joint CE and CSI feedback architecture deviates from traditional CE techniques, which estimate explicit downlink CSI. Instead, it employs multiple convolutional layers within the JEC module to extract channel feature information from the downlink CSI present in the received signals. Following this, the ``Reshape'' operation is applied to transform the feature matrix into a vector, and a fully connected (FC) layer compresses the high-dimensional feature vector into low-dimensional feature information $\mathbf{s} = \left[{s(1), \cdots, s(K)}\right]^{\mathrm{T}} \in \mathbb{C}^{K}$, as shown in Fig.~\ref{UJEFNet}. The JEC module corresponds to the encoder module $\mathcal{E}_{\alpha}(\cdot)$ discussed in Section~\ref{Section_CSI_FB}, with trainable parameter set $\alpha$, except that the source is now the received pilot signal, which contains the feature information of the downlink CSI, rather than the downlink CSI itself. The joint CE and CSI compression with DJSCC architecture is represented as Eq.~\eqref{encoder}.

As illustrated in the signal transmission procedure in Section~\ref{Section_CSI_FB}, the vector representing the feature information undergoes power normalization and OFDM mapping before being transmitted. At the BS side, the received signal at the $k$-th subcarrier can be expressed as:
\begin{equation}
    \label{Nr1_CSI_FB_receive}
    \mathbf{y}(k)={\mathbf{h}_{\mathrm{u}}}(k)s(k)+\mathbf{n}_{\mathrm{u}}(k),
\end{equation}
where ${\mathbf{h}_{\mathrm{u}}}(k) \in \mathbb{C}^{N_{\mathrm{BS}}}$ is uplink CSI of the $k$-th subcarrier, with the assumption that $N_{\mathrm{UE}}=1$.

At the BS, we use the maximum ratio combining (MRC) algorithm, denoted as $\mathcal{F}_{MRC}(\cdot)$ to instead the $\mathcal{F}(\cdot)$ in Eq.~\eqref{detection_algorithm}, to detect each transmitted signal at the occupied uplink subcarrier using the estimated spatial-frequency domain uplink CSI $\tilde{\mathbf{H}}_{\mathrm{u}} = [\tilde{\mathbf{h}}_{\mathrm{u}}(1), \cdots, \tilde{\mathbf{h}}_{\mathrm{u}}(K), \cdots, \tilde{\mathbf{h}}_{\mathrm{u}}(M)]^{\mathrm{T}} \in \mathbb{C}^{M \times N_{\mathrm{BS}}}$, as follows:
\begin{equation}
    \label{MRC}
    \hat{s}(k)=\mathcal{F}_{MRC}\left({\tilde{\mathbf{h}}_u(k), \mathbf{y}(k)}\right)=\frac{\tilde{\mathbf{h}}_{\mathrm{u}}(k)^\mathrm{H}}{\Vert{\tilde{\mathbf{h}}_{\mathrm{u}}(k)}\Vert_2}\mathbf{y}(k),
\end{equation}
where $k \in \{1, \cdots, K\}$ represents the $k$-th subcarrier for performing MRC detection on an individual basis.

It is worth emphasizing that, in $\tilde{\mathbf{H}}_{\mathrm{u}}$, the $k$-th row vector $\tilde{\mathbf{h}}_{\mathrm{u}}(k)$ corresponds to the first $k$-th element of the set $\tilde{\mathscr{H}}_{\mathrm{u}} = \{\tilde{\mathbf{h}}_{\mathrm{u}}(m)\}_{m=1}^M$ with $N_{\mathrm{UE}} = 1$.
% Only a subset of the $K$ subcarriers is utilized for CSI feedback, while the remaining $M - K$ subcarriers are allocated for the transmission of other control or data signals, which are not the focus of this paper.

After signal detection, the downlink CSI can be reconstructed based on the detected feature information $\hat{\mathbf{s}} = \left[\hat{s}(1), \cdots, \hat{s}(K)\right]^{\mathrm{T}}$ of the downlink channel using the CSI reconstruction module. The network architecture is shown in the lower left of Fig.~\ref{UJEFNet}. The CSI reconstruction module consists of multiple residual blocks for CSI reconstruction and transposed convolutional layers for up-sampling, utilizing CSI correlation in the frequency domain. The initially reconstructed downlink CSI can be denoted as $\overline{\mathbf{G}}_{\mathrm{d}} \in \mathbb{C}^{M \times N_{\mathrm{BS}}}$.

\subsection{Non-Ideal Uplink Estimation}
\label{Non-Ideal_Uplink_Estimation}
Unlike CSI feedback in most DJSCC architectures and uplink-assisted CSI feedback in SSCC architectures, we consider the case of non-ideal uplink CE. In the experimental section, i.e., Section \ref{section3}, we analyze its impact on system performance. This paper examines two CE schemes: traditional least squares (LS) estimation and a AI-based algorithm. For uplink CE with $N_r = 1$, we set $L = 1$, define the uplink pilot interval as $g_\mathrm{u}$, and set the pilot value to $1$. To differentiate between the lengths of the uplink and downlink frequency-domain pilot signals, we use $M_p$ to denote the number of subcarriers occupied by the uplink pilot signal, where $M_p$ satisfies $M_p = \lceil \frac{M}{g_\mathrm{u}} \rceil$.

Using the traditional LS estimation algorithm and assuming that the uplink pilots consist of 1, the estimated channel matrix at the pilot positions is given by:
\begin{equation}
    \label{LS_CE_UL}
    \tilde{\dot{\mathbf{H}}}_{\mathrm{u}}=\mathbf{T}^{\mathrm{T}}\mathbf{H}_{\mathrm{u}}+\mathbf{N}_{\mathrm{u}}
    % =\dot{\mathbf{H}}_\mathrm{u}+\mathbf{N}_{\mathrm{u}}
    ,
\end{equation}
where $\tilde{\dot{\mathbf{H}}}_{\mathrm{u}} \in \mathbb{C}^{M_p \times N_{\mathrm{BS}}}$ and $\mathbf{H}_{\mathrm{u}}=\left[{\mathbf{h}_{\mathrm{u}}(1),\cdots,\mathbf{h}_{\mathrm{u}}(M)}\right]^{\mathrm{T}} \in \mathbb{C}^{M \times N_{\mathrm{BS}}}$ represent the estimated uplink channel and the actual uplink channel, respectively. Additionally, $\mathbf{N}_{\mathrm{u}} \in \mathbb{C}^{M_p \times N_{\mathrm{BS}}}$ denotes the AWGN in the uplink channel with the same distribution of uplink noise in Eq.~\eqref{Nr1_CSI_FB_receive}.

After performing LS-based CE at the pilot positions, linear interpolation is applied to estimate the CSI at non-pilot subcarriers, which is given by:
\begin{equation}
    \label{LS_interpolation}
    \tilde{{\mathbf{H}}}_{\mathrm{u}}=\mathcal{F}_{\mathrm{line}}\left({\tilde{\dot{\mathbf{H}}}_{\mathrm{u}}}\right) \in \mathbb{C}^{M \times N_{\mathrm{BS}}}.
\end{equation}

\begin{figure}[!t]
	\centering
	\includegraphics[width=3.4in]{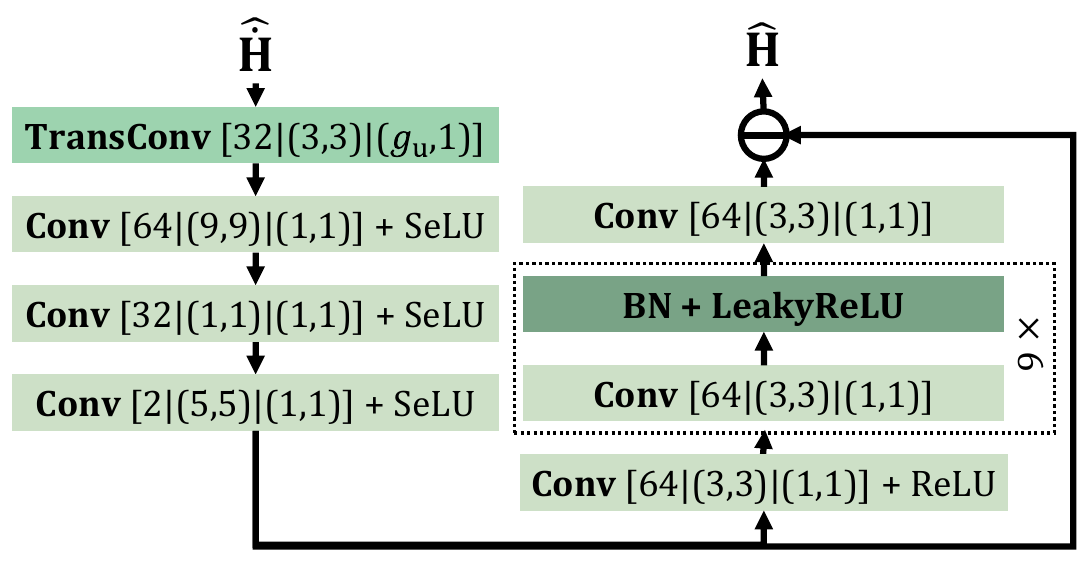}
	\caption{Architecture of the AI-based uplink CE network framework. ``SeLU'' in the figure refers to the Scaled Exponential Linear Unit (SeLU) activation function.}
	\label{CE_Framework}
\end{figure}
For the AI-based CE algorithm, we leverage AI to directly capture the frequency-domain correlation of the channel and environmental information, as shown in Fig.~\ref{CE_Framework}. This enables interpolation and extraction of channel characteristics from the training dataset, thereby enhancing the reconstruction accuracy of the uplink channel, which can be expressed as:
\begin{equation}
    \label{AI_CE}
    \tilde{{\mathbf{H}}}_{\mathrm{u}}=\mathcal{F}_{\mathrm{AI}}\left({\tilde{\dot{\mathbf{H}}}_{\mathrm{u}}}\right).
\end{equation}

It is important to note that, for simplicity, we uniformly use $\tilde{{\mathbf{H}}}_{\mathrm{u}}$ to denote the estimated uplink channel matrix in Eq.~\eqref{LS_interpolation} and Eq.~\eqref{AI_CE}. However, in practice, for the same $\tilde{\dot{\mathbf{H}}}_{\mathrm{u}}$, these two estimates are not identical.

\subsection{Uplink Assisted CSI Refine}
In FDD systems, despite the uplink and downlink channels working at different frequencies, there exists some correlation between their propagation environments when the duplex distance is not excessively large \cite{Spatial_reciprocity}. This correlation arises from shared scatterer environments and corresponding angles of departure and arrival. Many existing studies leverage the partial correlation between the uplink and downlink for network design \cite{Bi-Directional, CA_Net,DNN_TypeII, GMM_Codebook}.

Given that the reciprocity of the uplink and downlink channels is captured in the angular domain, we propose a ``\emph{Joint Refine}'' module to enhance the accuracy of CSI reconstruction by leveraging uplink CSI. At the BS, we jointly design three modules—channel decoding, CSI reconstruction, and ``\emph{Joint Refine}''—which operate in the angular-frequency domain. The downlink CSI dimension is reconstructed using the CSI reconstruction module after the BS receives the compressed pilot matrix containing the feature information of the downlink CSI. This reconstruction result can be regarded as the initially reconstructed angular-frequency domain CSI $\overline {\mathbf{G}}_{\mathrm{d}}$.

The uplink spatial-frequency domain CSI $\tilde {\mathbf{H}}_{\mathrm{u}}$, obtained through uplink CE, first needs to be transformed to the uplink CSI matrix in the angular-frequency domain, denoted as $\hat {\mathbf{G}}_{\mathrm{u}}=\tilde {\mathbf{H}}_{\mathrm{u}}\mathbf{F}$.

To utilize the reciprocity of the uplink channel for enhance the reconstruction accuracy of downlink CSI, the matrices $\hat {\mathbf{G}}_{\mathrm{u}}$ and $\overline {\mathbf{G}}_{\mathrm{d}}$ are jointly sent to the ``\emph{Joint Refine}'' module to reconstruct finial downlink CSI $\hat {\mathbf{G}}_{\mathrm{d}} \in \mathbb{C}^{M \times N_{\mathrm{BS}}}$, as shown in the upper right of Fig.~\ref{UJEFNet}. Notably, the ``\emph{Joint Refine}'' module comprises $T$ ``Refine Blocks'', where the inputs of each ``Refine Block'' include both the output of the previous layer and the uplink CSI as auxiliary information. These inputs are firstly concatenated in the $\mathrm{Concat}$ layer. The one of the inputs in the first ``\emph{Joint Refine}'', denoted as $\bar{\mathbf{G}}_{\mathrm{d}}^{(0)}$, corresponds to the output of the CSI reconstruction module, $\bar{\mathbf{G}}_{\mathrm{d}}$.

The CSI reconstruction module and the ``\emph{Joint Refine}'' module correspond to the decoder module $\mathcal{D}_{\beta}(\cdot)$ in Section~\ref{Section_CSI_FB}. The reconstructed downlink CSI in the angular-frequency domain can be expressed as:
\begin{equation}
    \label{Rec_CSI}
    \hat {\mathbf{G}}_{\mathrm{d}}=\mathcal{D}_{\beta}\left({\hat{\mathbf{s}}, \hat {\mathbf{G}}_{\mathrm{u}}}\right).
\end{equation}
\begin{table*}[!t]
\renewcommand{\arraystretch}{1.15} % 增加行距至1.3倍
\caption{Network Architecture Configuration}
\label{net_config}
\centering
\begin{tabular}{c|c|c|c|c}
\toprule
\textbf{Module Name} & \textbf{Layer Name (Sequential)} & \textbf{Parameters Configuration} & \textbf{Input Shape} & \textbf{Output Shape} \\ 
\midrule
\multirow{3}{*}{\textbf{JMP module}} & Conv+BN $\times{2}$& 2\textbar(7,7)\textbar(1,1) & $N_p \times N_t \times 2$ & $N_p \times N_t \times 2$ \\ 
 % & Conv+BN & 2\textbar(7,7)\textbar(1,1) & $N_p \times N_t \times 2$ & $N_p \times N_t \times 2$ \\ 
 & Reshape & / & $N_p \times N_t \times 2$ & $2N_pN_t$ \\ 
 & Dense & $2K$ & $2N_pN_t$ & $2K$ \\ 
 \cmidrule{1-5}
\multirow{7}{*}{\textbf{CSI Reconstruction}} & Dense & 2048 & $2K$ & 2048 \\ 
 & Reshape & / & 2048 & $32 \times 32 \times 2$ \\ 
 & Conv+BN & 2\textbar(7,7)\textbar(1,1) & $32 \times 32 \times 2$ & $32 \times 32 \times 2$ \\ 
 & Residual Block $\times5$ & As shown in Fig.~\ref{UJEFNet} & $32 \times 32 \times 2$ & $32 \times 32 \times 2$ \\ 
 % & Residual Block & As shown in Fig.~\ref{UJEFNet} & $32 \times 32 \times 2$ & $32 \times 32 \times 2$ \\ \cline{2-5}
 % & Residual Block & As shown in Fig.~\ref{UJEFNet} & $32 \times 32 \times 2$ & $32 \times 32 \times 2$ \\ \cline{2-5}
 % & Residual Block & As shown in Fig.~\ref{UJEFNet} & $32 \times 32 \times 2$ & $32 \times 32 \times 2$ \\ \cline{2-5}
 % & Residual Block & As shown in Fig.~\ref{UJEFNet} & $32 \times 32 \times 2$ & $32 \times 32 \times 2$ \\ \cline{2-5}
 & TransConv+BN  & 16\textbar(3,3)\textbar(2,1) & $32 \times 32 \times 2$ & $64 \times 32 \times 16$ \\ 
 & TransConv+BN & 16\textbar(3,3)\textbar(2,1) & $64 \times 32 \times 16$ & $128 \times 32 \times 16$ \\ 
 & TransConv+BN & 2\textbar(3,3)\textbar(2,1) & $128 \times 32 \times 16$ & $256 \times 32 \times 2$ \\ 
 \bottomrule
\end{tabular}
\end{table*}
\subsection{End-to-end Network Architecture}
\label{end_to_end_manner}
Joint training of downlink pilot design, downlink CE, and downlink CSI compression enhances information compression efficiency. The end-to-end design of a fully AI-driven network enables the UE to infer feature information of the downlink CSI directly from the received pilot signals. With the same feedback overhead, extracting feature information directly from lower-dimensional received pilots and compressing them yields an equivalent reduction in the compression rate. Consequently, this joint design improves CSI reconstruction accuracy.

Simultaneously considering downlink pilot design, downlink CE, and uplink CSI-assisted downlink CSI reconstruction allows the network to allocate more attention to the distinctive components of downlink and uplink CSI during downlink CE and feature extraction, rather than focusing on reciprocal parts that uplink CSI can supplement. This approach leverages uplink CSI to complement reciprocal components, ultimately enhancing CSI reconstruction accuracy. Additionally, joint training of the CSI reconstruction module with the ``\emph{Joint Refine}'' module further contributes to this improvement.

During training, the network accounts for inaccuracies in downlink CE, uplink CE, and CSI feedback, adjusting its parameters accordingly to mitigate their impact on overall performance. Additionally, DJSCC prevents performance degradation caused by the ``cliff effect.'' This approach enhances end-to-end performance and makes the system more suitable for deployment in non-ideal real-world scenarios.

The whole network is trained in an end-to-end manner with the parameter set $\Theta=\{\alpha,\beta,\mathcal{Q}\}$ to prevent performance degradation caused by distribution bias, and the parameter optimization space of different modules is optimized with the same optimization objective, as:
\begin{equation}
    \label{Loss_UJEFNet}
    \mathop {{\rm{minimize}}}\limits_{\Theta} {\left\Vert {f\left({\hat{\mathbf{s}}, \hat {\mathbf{G}}_{\mathrm{u}};\Theta}\right)-\mathbf{G}_{\mathrm{d}}} \right\Vert_{\mathrm{F}}^2},
\end{equation}
where $\mathbf{G}_{\mathrm{d}}$ is downlink CSI in the angular-frequency domain, which satisfies DFT transform with spatial-frequency domain CSI, i.e., $\mathbf{G}_{\mathrm{d}}=\mathbf{H}_{\mathrm{d}}\mathbf{F}$.

% During training, the input distribution of subsequent modules matches the output distribution of preceding modules, preventing performance degradation caused by distribution mismatch. This approach stabilizes end-to-end network performance and facilitates the attainment of an optimal solution. 

\section{Experimental results}
\label{section3}

The UJEFNet proposed in this paper presents a generalized architecture for joint pilot design, CE, and CSI feedback with uplink channel assistance. In this section, we first describe the simulation scenario configuration and dataset generation process. Using the generated dataset, we then test the correlation between uplink and downlink channels in the spatial and angular domains. Next, we evaluate the effectiveness of the proposed network and perform ablation experiments to assess the contribution of different modules to overall performance. We also evaluate the impact of non-ideal uplink CE on the CSI feedback task under the DJSCC architecture. Finally, we examine and discuss the scalability of the proposed network

Each sample pair, consisting of an uplink CSI matrix and a downlink CSI matrix, is generated according to the 3rd Generation Partnership Project (3GPP) TR 38.901 \cite{3GPP38901} using Sionna \cite{Sionna}, a Python-based open-source library. Sionna has been widely used in AI-based communication system research, including studies such as \cite{Sionna_GPT, Sionna_receiver}. The center frequencies of the uplink and downlink channels are set to 5.6 GHz and 5.4 GHz, respectively. The channel scenario follows a clustered delay line (CDL) model, specifically CDL-C. Single-polarization antenna arrays, compliant with 3GPP TR 38.901, are deployed at both the UE and BS, with the number of antennas set to $N_{\mathrm{UE}}=1$ and $N_{\mathrm{BS}}=32$, respectively. The channel delay spread is set to 100 ns. The number of subcarriers $M$ for both uplink and downlink is 256. The initial pilot intervals for uplink and downlink CE, $g_\mathrm{u}$ and $g_{\mathrm{d}}$, are set to 1 and 4, respectively, meaning the number of pilot-occupied subcarriers is $M_p=256$ and $N_p=64$. The occupied OFDM symbols for uplink and downlink CE collectively represented as $L$ in Section \ref{section1}, are set to 1 and 16, respectively. The training, validation, and test sets, generated using the Sionna simulator, contain 100,000, 30,000, and 10,000 sample pairs, respectively. The initial learning rate is set to 0.001. If the validation loss does not decrease for more than 20 consecutive epochs, the learning rate is reduced by half. The Adam optimizer is employed for optimization. The batch size is set to 200, and the model is trained for 200 epochs. The network architecture is implemented using TensorFlow and Keras. With this configuration, the specific implementation parameters of each module are provided in Table~\ref{net_config}. Other activation functions and network configurations are illustrated in Fig.~\ref{UJEFNet}.
% In the ``$\mathrm{Reshape}$'' layer, we use ``$\rightarrow$'' to indicate the transformation from the original to the new dimension.

\subsection{Reciprocity Testing}

\begin{figure}[!t]
	\centering
	\includegraphics[width=3.4in]{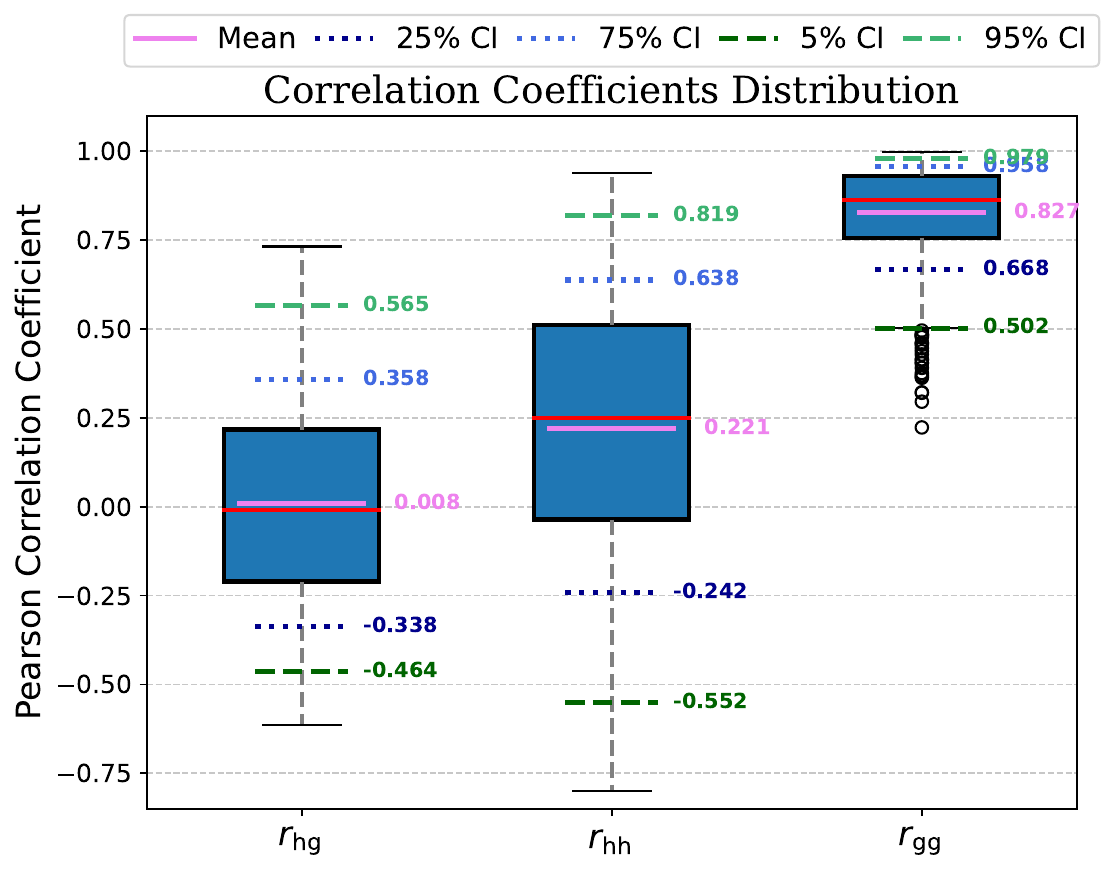}
	\caption{The confidence intervals for Pearson correlation coefficients with $r_{\mathrm{hg}}$, $r_{\mathrm{hh}}$, $r_{\mathrm{gg}}$.}
	\label{CI}
\end{figure}

We sampled the first subcarrier of the generated CSI samples to evaluate the correlation between uplink and downlink CSI in the angular and spatial domains. The correlation strength is measured using the Pearson correlation coefficient \cite{Bi-Directional}, which is defined as
\begin{equation}
    \label{Pearson}
    r\left({\mathbf{a}, \mathbf{b}}\right) = \frac{\sum_{i=1}^N\left(a_i-\bar{a}\right)\left(b_i-\bar{b}\right)}{\sqrt{\sum_{i=1}^N\left(a_i-\bar{a}\right)^2}\sqrt{\sum_{i=1}^N\left(b_i-\bar{b}\right)^2}}.
\end{equation}
$r\left({\mathbf{a}, \mathbf{b}}\right)$ represents the correlation between the vectors $\mathbf{a} \in \mathbb{R}^{N}$ and $\mathbf{b} \in \mathbb{R}^{N}$. The elements $a_i$ and $b_i$ denote the $i$-th element of $\mathbf{a}$ and $\mathbf{b}$, respectively. A value of $r\left({\mathbf{a}, \mathbf{b}}\right)$ close to 1 indicates a strong positive correlation, while a value close to -1 signifies a strong negative correlation. Conversely, a value near 0 suggests a weak or no correlation. According to \cite{Bi-Directional, CA_Net}, the partial reciprocity between uplink and downlink in the FDD system is reflected in the modulus of the CSI, so we separately evaluate the following metrics
\begin{itemize}
\item[$\bullet$]The Pearson correlation coefficients of uplink CSI in the angular domain and downlink CSI in the spatial domain can be defined as $r_{\mathrm{hg}}=r\left({\left\vert \mathbf{h}_{\mathrm{u}}(1) \right\vert^2, \left\vert \mathbf{g}_{\mathrm{d}}(1) \right\vert^2}\right)$.
\item[$\bullet$]The Pearson correlation coefficients of uplink CSI and downlink CSI both in the spatial domain can be defined as $r_{\mathrm{hh}}=r\left({\left\vert \mathbf{h}_{\mathrm{u}}(1) \right\vert^2, \left\vert \mathbf{h}_{\mathrm{d}}(1) \right\vert^2}\right)$.
\item[$\bullet$]The Pearson correlation coefficients of uplink CSI and downlink CSI both in the angular domain can be defined as $r_{\mathrm{gg}}=r\left({\left\vert \mathbf{g}_{\mathrm{u}}(1) \right\vert^2, \left\vert \mathbf{g}_{\mathrm{d}}(1) \right\vert^2}\right)$.
\end{itemize}

We randomly selected 1,000 pairs of datasets from the generated datasets to compute the values of the three aforementioned metrics. We then plotted a box plot of the confidence intervals for the Pearson correlation coefficients. The experimental results are presented in Fig.~\ref{CI}.

From the experimental results in Fig.~\ref{CI}, we observe that the generated uplink and downlink datasets exhibit a strong correlation in the angular frequency domain and a weak correlation in the spatial frequency domain, as indicated by higher and lower Pearson correlation coefficients, respectively. Moreover, the mean Pearson correlation coefficient between the different domains is approximately zero, indicating that there is no significant correlation. Therefore, we first apply the DFT to convert the acquired uplink CSI from the spatial frequency domain CSI to the angular frequency domain. Then, we utilize the uplink CSI in the angular frequency domain to assist in reconstructing the downlink angular frequency domain CSI. Furthermore, the entire downlink CE and feedback process is designed and evaluated in the angular frequency domain to maximally utilize the reciprocity.

\subsection{Performance Impact Analysis of Uplink Channel Estimation}
\label{EX_Non_Ideal_CE}
\begin{figure}[!t]
	\centering
	\includegraphics[width=3.4in]{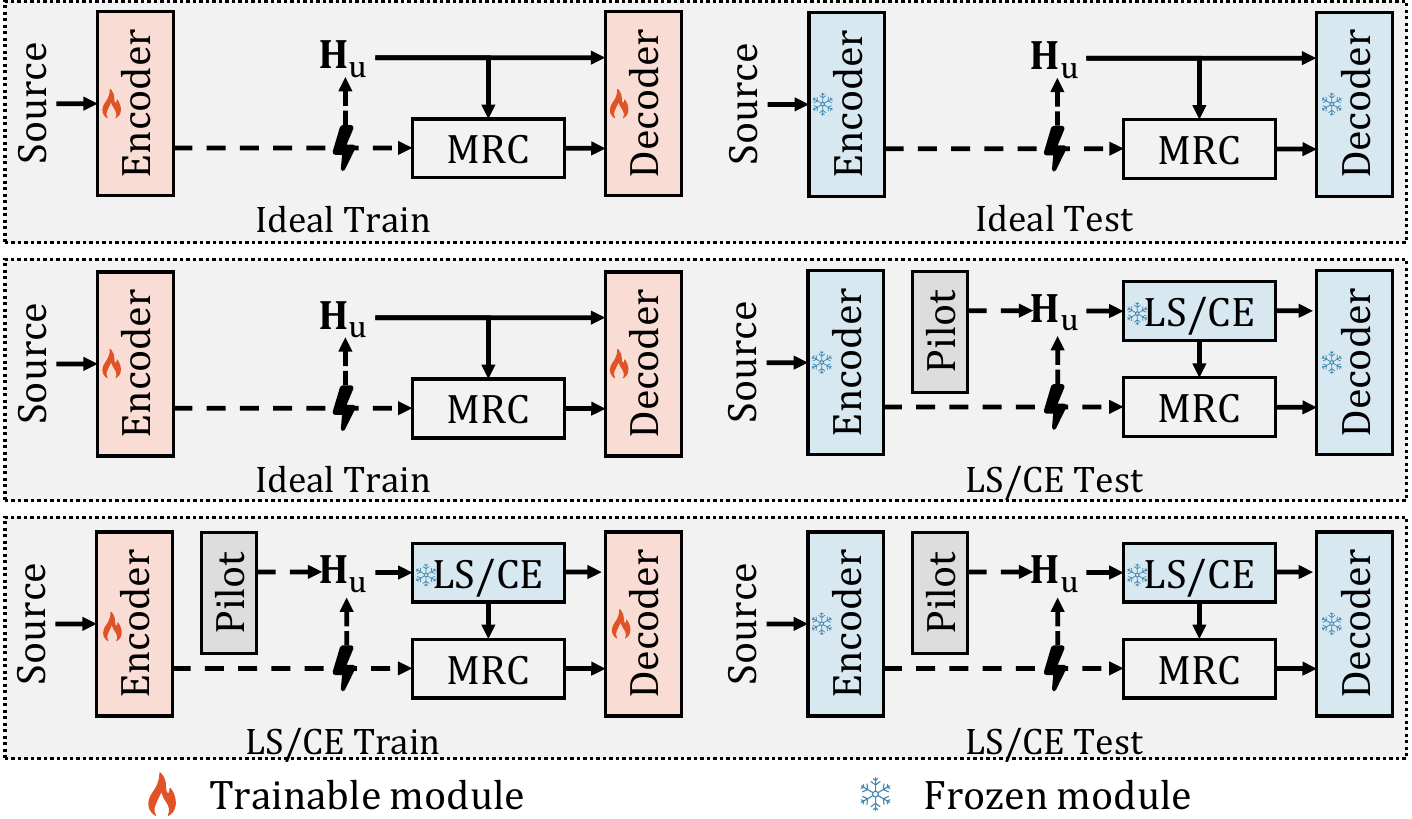}
	\caption{We outline the training and testing strategies for UJEFNet under non-ideal uplink CE. The networks at the transmitter and receiver are simplified using ``Encoder'' and ``Decoder'' representations, respectively. To indicate parameter updates during training, we use the fire flag, signifying that the corresponding module parameters are trainable at that stage. Conversely, the snowflake flag is used to denote modules with frozen parameters.}
	\label{Non_Ideal_CE_strategy}
\end{figure}
In this paper, we first investigate the impact of uplink CE errors on the performance of the DJSCC architecture. Based on the proposed UJEFNet architecture, we evaluate three network training and testing strategies, as illustrated in Fig.~\ref{Non_Ideal_CE_strategy}. The first strategy assumes that the network is trained and tested under ideal uplink CE, where the uplink channel is directly fed into the MRC receiver and decoder for signal detection and assisting downlink CSI reconstruction, respectively. This scheme serves as an upper performance bound, as it represents an ideal assumption, yet practically unattainable. The second strategy corresponds to most DJSCC architecture \cite{ADJSCC, JFPNet, JEFPNet}, where the network is trained under ideal uplink CE but practically tested with non-ideal estimation, such as LS estimation or AI-based CE methods (denoted as $\mathrm{CE}$ in Fig.~\ref{Non_Ideal_CE_strategy}), thereby ignoring the impact of CE errors. The third strategy introduces uplink CE errors during the training phase through an end-to-end learning approach, allowing the decoder to adapt to the distribution bias caused by these errors. The same CE algorithm is applied during both training and testing, enabling UJEFNet to achieve more robust performance under practical conditions.

In this experiment, we assume that the downlink CE SNR is set to $\mathrm{SNR_{ce}}=1/\sigma_{\mathrm{d}}^2 = 10$ dB, and the feedback bandwidth $K = 16$. We vary the uplink feedback channel SNR, denoted as $\mathrm{SNR_u}=1/\sigma_{\mathrm{u}}^2$, and adopt different network training and testing strategies to evaluate the impact of uplink CE errors on the network performance by comparing the NMSE of downlink CSI reconstruction. The evaluation metric NMSE is defined as:
\begin{equation}
    \label{NMSE}
    \mathrm{NMSE}=\mathbb{E}\left\{{\frac{\Vert{\mathbf{G}_{\mathrm{d}}-\hat{\mathbf{G}}_{\mathrm{d}}}\Vert_\mathrm{F}^2}{\Vert{\mathbf{G}_{\mathrm{d}}}\Vert_{\mathrm{F}}^{2}}}\right\}.
\end{equation}

\begin{figure}[!t]
	\centering
	\includegraphics[width=3.4in]{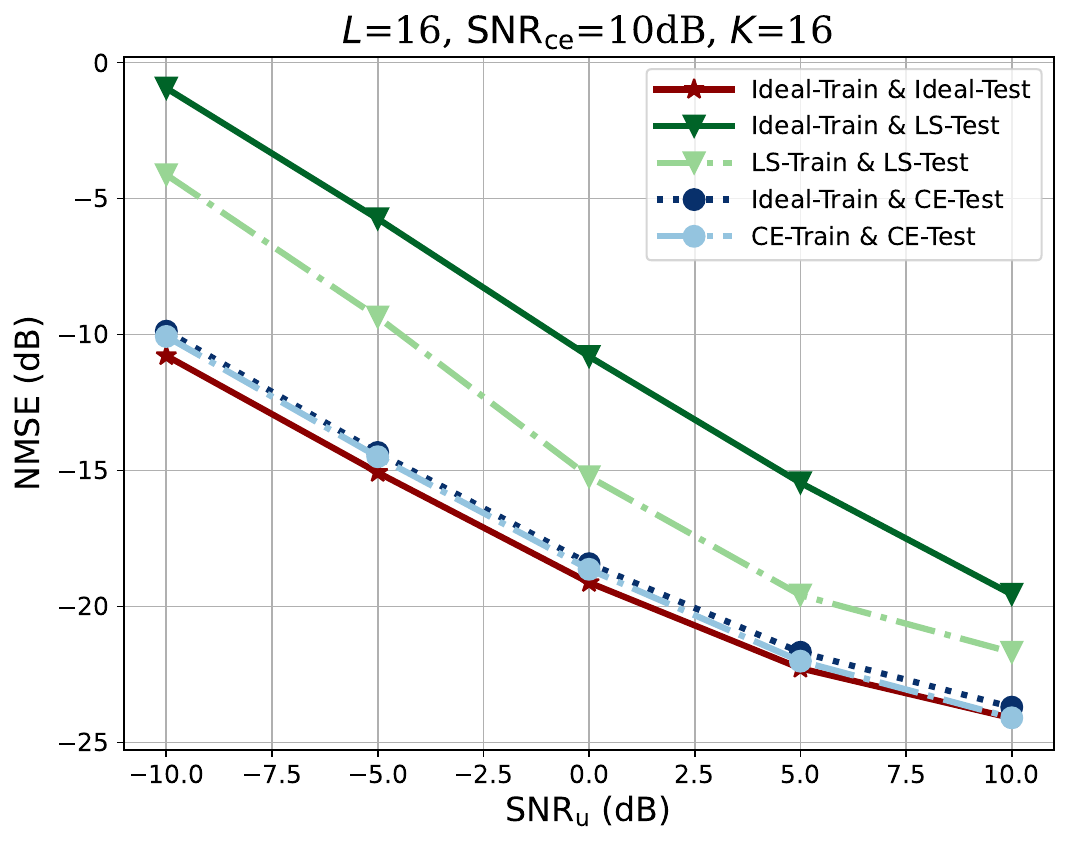}
	\caption{Performance of UJEFNet under different training and testing strategies and various CE algorithms in non-ideal uplink CE.}
	\label{Non_Ideal_DiffCase}
\end{figure}
From the experimental results in Fig.~\ref{Non_Ideal_DiffCase}, we observe that the ideally trained UJEFNet network experiences significant performance degradation when subjected to uplink CE errors during testing, particularly with the LS algorithm, which exhibits low CE accuracy. This degradation occurs because the uplink CE error leads to inconsistency between the received feature information and the original feature information. As the CE error increases, the accuracy of the MRC detector decreases accordingly. In contrast, the AI-based CE algorithm achieves a higher accuracy than the LS algorithm, resulting in a smaller performance loss. Specifically, compared to the upper bound, the performance loss for training under ideal CE and testing under LS estimation reaches 9.84 dB at $\mathrm{SNR_{\mathrm{u}}} = -10$ dB, while the performance loss using the AI-based CE algorithm with the same training and testing strategy is only 0.90 dB. However, it should be noted that the AI-based CE algorithm is based on LS estimation, which introduces additional computational complexity, model storage overhead, and inference delay compared to the LS algorithm. Although these overheads are manageable for BS, the latency introduced by the AI model is unfavorable for latency-sensitive scenarios, such as high-speed railway communication and autonomous driving.

By introducing CE errors during the training phase in an end-to-end training manner, the decoder can effectively adapt to the input distribution bias caused by CE errors, thereby enhancing the network's robustness to uplink CE errors and improving the accuracy of downlink CSI reconstruction. As shown in Fig.~\ref{Non_Ideal_DiffCase}, for the LS CE algorithm, incorporating CE errors during training provides a performance gain of 3.18 dB at $\mathrm{SNR_{\mathrm{u}}} = -10$ dB and 4.42 dB at $\mathrm{SNR_{\mathrm{u}}} = 0$ dB, compared to the strategy trained under ideal conditions. Similarly, for the AI-based CE algorithm, the same strategy achieves a performance gain of 0.19 dB at $\mathrm{SNR_{\mathrm{u}}} = -10$ dB and 0.39 dB at $\mathrm{SNR_{\mathrm{u}}} = 10$ dB. In particular, at $\mathrm{SNR_{\mathrm{u}}} = 10$ dB, the performance is only 0.013 dB away from the upper bound, which can often be ignored in practical scenarios.

\begin{figure}[!t]
	\centering
	\includegraphics[width=3.4in]{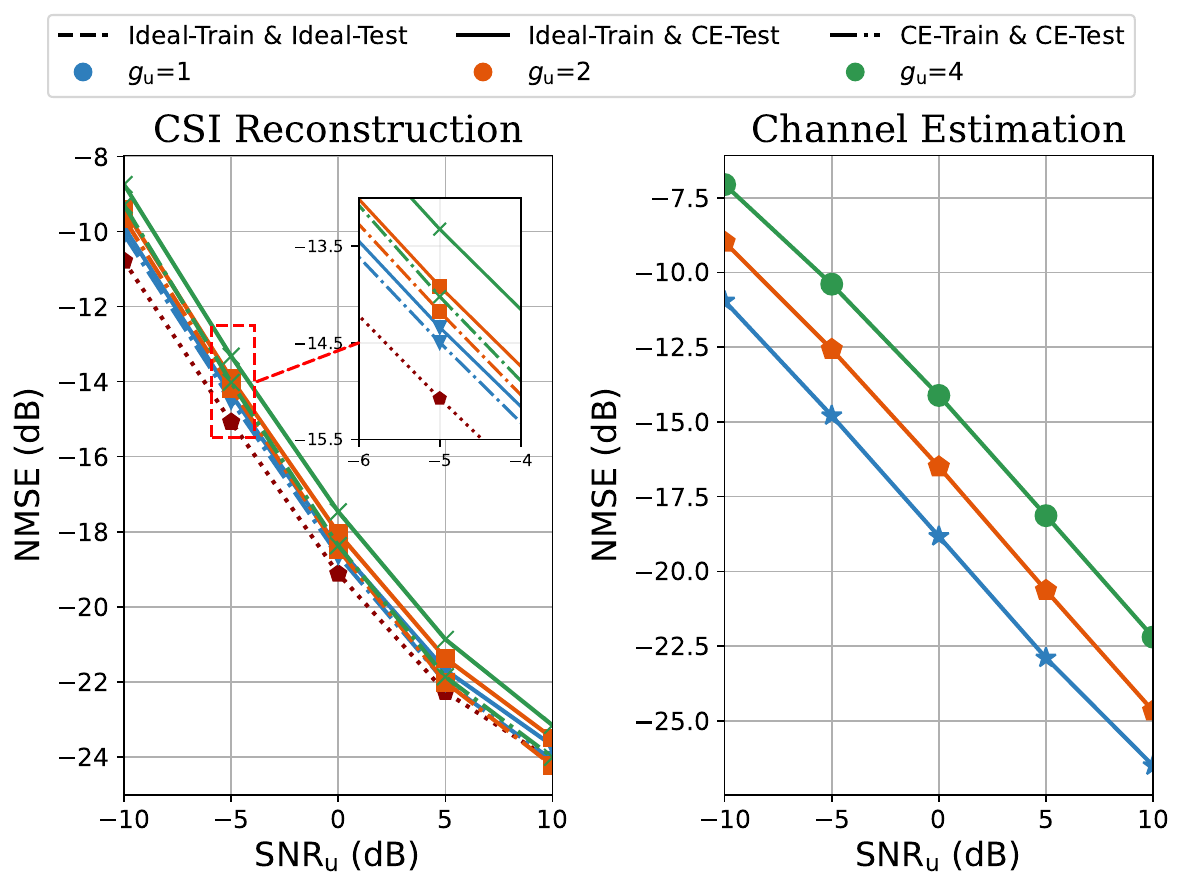}
	\caption{Performance of UJEFNet under different training and testing strategies and pilot intervals with AI-based CE in non-ideal uplink CE. We use color coding to distinguish pilot intervals $g_{\mathrm{u}}$ and employ a line-style differentiation strategy for clarity.}
	\label{Non_Ideal_DiffG}
\end{figure}
We further evaluate the third strategy of AI-based CE under different uplink pilot intervals, where the pilot interval is set as $g_\mathrm{u} \in \{1, 2, 4\}$. Other settings remain unchanged. The experimental results are shown in Fig.~\ref{Non_Ideal_DiffG}, where the right subfigure presents the CE accuracy, and the left subfigure shows the CSI reconstruction accuracy. As depicted in the right figure, the CE error increases as the pilot interval becomes larger. Specifically, compared to $g_\mathrm{u} = 1$, the performance loss is approximately 2 dB under $g_\mathrm{u} = 2$ and around 4 dB to 5 dB under $g_\mathrm{u} = 4$ for $\mathrm{SNR_\mathrm{u}} \in [-10 \text{ dB}, 10 \text{ dB}]$. From the left subfigure, we observe that as the CE error increases, the impact on CSI reconstruction under the second strategy becomes more significant. For instance, at $\mathrm{SNR_\mathrm{u}} = -5$ dB, the performance loss relative to the upper bound is 0.74 dB for $g_\mathrm{u} = 1$, 1.16 dB for $g_\mathrm{u} = 2$, and 1.75 dB for $g_\mathrm{u} = 4$. This degradation stems from the fact that larger CE errors introduce more deviation between the decoder input and the original input distribution during training. However, consistent with previous findings, the third strategy, which introduces CE errors during training, consistently outperforms the second strategy at all pilot intervals, with the performance gain increasing as the pilot interval grows. Specifically, at $\mathrm{SNR_\mathrm{u}} = -5$ dB, the performance gain is 0.16 dB for $g_\mathrm{u} = 1$, 0.26 dB for $g_\mathrm{u} = 2$, and 0.70 dB for $g_\mathrm{u} = 4$.

\subsection{Analysis of the Effectiveness of the Proposed Network}
\label{EX_Ideal_CE}
The experimental results in Section \ref{EX_Non_Ideal_CE} demonstrate that the impact of uplink CE errors on CSI reconstruction can be effectively mitigated through appropriate CE algorithms and reasonable network training strategies, albeit at the cost of additional computational and storage resources at the BS. Therefore, in the subsequent experiments, to analyze the performance of the proposed network and verify the effectiveness of the proposed architecture more concisely and clearly, we assume an ideal uplink CE.

\begin{figure*}[!t]
  \centering
  \subfloat[]{
    \label{Ideal_Performance_K16}
    \includegraphics[width=3.4in]{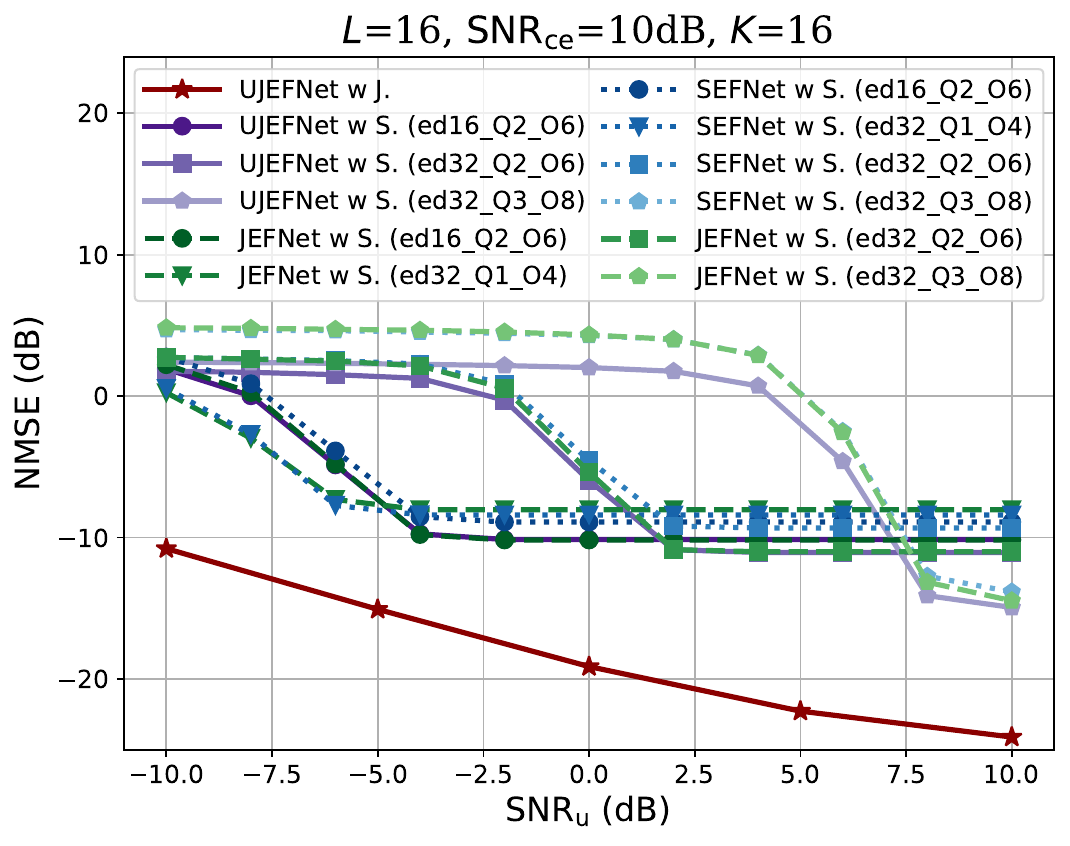}
    
  }
  \subfloat[]{
    \label{Ideal_Performance_K8}
    \includegraphics[width=3.4in]{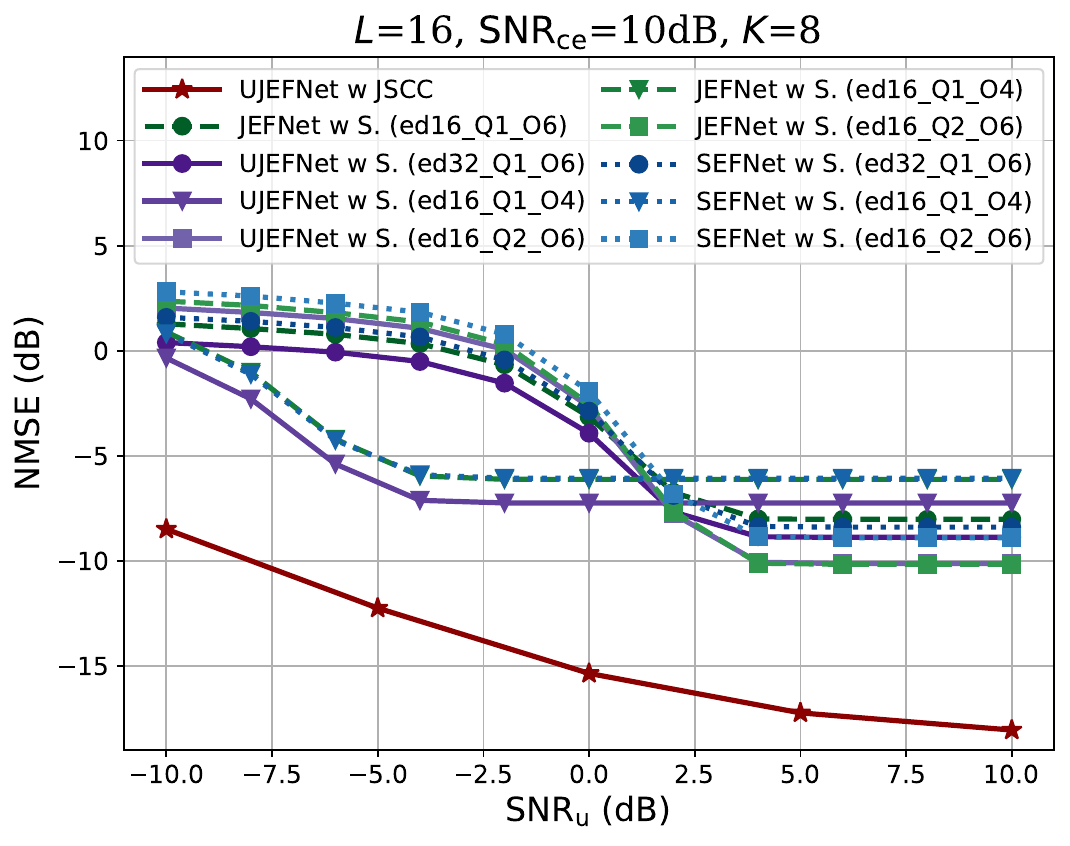}
    
  }\\
  \caption{Performance comparison of UJEFNet under the DJSCC architecture (denoted as ``w J.'') and various network architectures under the SSCC architecture (denoted as ``w S.'') with different feedback bandwidths, i.e., \protect\subref{Ideal_Performance_K16} $K=16$ and \protect\subref{Ideal_Performance_K8} $K=8$. ``ed16\_Q2\_O6'' indicates that the curve corresponds to a UE-side CSI compression module with 16 output neurons, a quantization bit number of 2, and 256QAM modulation.}
  \label{Ideal_Performance_Diff_K}
\end{figure*}

In this subsection, we aim to validate the effectiveness of the proposed UJEFNet based on the DJSCC architecture through simulation experiments. We set the downlink CE overhead to $L = 16$ and the downlink CE $\mathrm{SNR_{ce}} = 10$ dB. For the comparison algorithms, we consider UJEFNet, JEFNet, and SEFNet under the SSCC architecture. The main structure of UJEFNet under the SSCC architecture is similar to that of UJEFNet under the DJSCC architecture, with the feature information transfer phase replaced by transmission under the SSCC architecture. For JEFNet and SEFNet, we adopt the architecture shown in Fig.~\subref*{SEFNet} and Fig.~\subref*{JEFNet_SSCC}, respectively.

For the network under the SSCC architecture, we adopt the quantization, channel coding, modulation, demodulation, channel decoding, and de-quantization operations, same as \cite{ADJSCC}, to achieve feature information of downlink CSI transmission. Additionally, we also introduce the offset module from \cite{CsiNet_Plus, ADJSCC} to compensate for quantization errors. To ensure a fair comparison, we match the communication bandwidth $K$ between the SSCC and DJSCC architectures by adjusting the number of neurons $e$ in the encoder's output layer, the quantization bits $B$, the modulation order $O$, and the code rate $R$ in the SSCC architecture. The relationship between them can be expressed as $K={eB}/{RO}$. We evaluate the performance of multiple network architectures with bandwidths $K$ of 16 and 8. The experimental results are presented in Fig.~\subref*{Ideal_Performance_K16} and Fig.~\subref*{Ideal_Performance_K8}, respectively.

From the experimental results in Fig.~\subref*{Ideal_Performance_K16} and Fig.~\subref*{Ideal_Performance_K8}, it can be observed that the network under the SSCC architecture suffers from a severe ``cliff effect,'' i.e., when the $\mathrm{SNR_u}$ of the feedback channel drops below the processing capability of the channel coding, the CSI reconstruction accuracy drastically decreases to an unusable level. A comparison between the ``UJEFNet w S. (ed16\_Q2\_O6)'' and ``UJEFNet w S. (ed32\_Q2\_O6)'' curves at $K = 16$ reveals that at high SNRs, e.g., $\mathrm{SNR_u} = 10$ dB, UJEFNet with $e = 32$ outperforms UJEFNet with $e = 16$ by approximately 0.9 dB. However, the former experiences the cliff effect earlier as $\mathrm{SNR_u}$ decreases. This phenomenon occurs because a larger number of encoder output neurons $e$ leads to a lower compression rate and higher feature information density, resulting in improved reconstruction accuracy when the decoder successfully decodes the received signal. Nevertheless, with the same $K$ and quantization bits $B$, the two curves correspond to code rates $R$ of 1/3 and 2/3, respectively. The ``UJEFNet w S. (ed16\_Q2\_O6)'' has more channel coding redundancy, which provides better resistance to channel noise. Comparing the ``UJEFNet w S. (ed32\_Q3\_O8)'' curve with the ``UJEFNet w S. (ed32\_Q2\_O6)'' curve, it can be observed that UJEFNet with $Q = 3$ achieves better performance than UJEFNet with $Q = 2$ at high SNR, e.g., $\mathrm{SNR_u} = 10$ dB. However, it experiences the ``cliff effect'' earlier. This is because a higher number of quantization bits and modulation order reduces quantization error and preserves more feature information. Nevertheless, with the same $K$, the code rates of the two curves are relatively close, i.e., 2/3 and 3/4, respectively. However, ``UJEFNet w S. (ed32\_Q3\_O8)'' adopts a higher modulation order, which is more prone to symbol misclassification and makes the network more sensitive to noise.

Suffering the ``cliff effect'' will lead to additional retransmission overhead and delay. Fortunately, the proposed UJEFNet under the DSCC architecture effectively avoids this issue. As $\mathrm{SNR_u}$ decreases, the accuracy of the CSI reconstruction gradually degrades and consistently outperforms UJEFNet under the SSCC architecture. Although both architectures share the same main backbone network, UJEFNet under the DJSCC architecture leverages end-to-end training to combat channel fading better and protect critical feature information, thereby achieving superior CSI reconstruction performance. Specifically, within the range of $\mathrm{SNR_u} \in [-10 \text{ dB}, 10 \text{ dB}]$, UJEFNet under the DJSCC architecture achieves a performance gain of 9.2 dB to 14.0 dB compared to UJEFNet under the SSCC architecture.

\subsection{Ablation Experiments}
\begin{figure}[!t]
	\centering
	\includegraphics[width=3.4in]{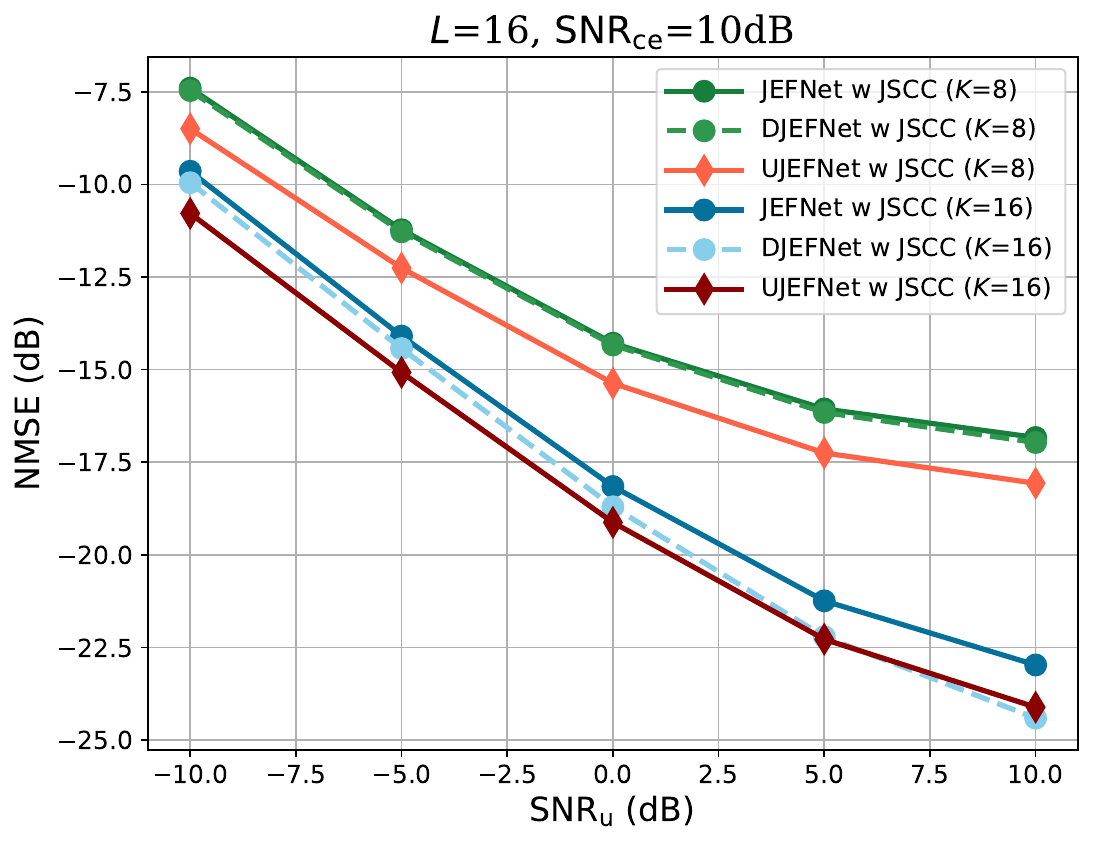}
	\caption{Ablation experiments: Evaluating the performance gain provided by uplink CSI assistance to the network.}
	\label{Ideal_Ablation}
\end{figure}
\begin{figure}[!t]
	\centering
	\includegraphics[width=3.4in]{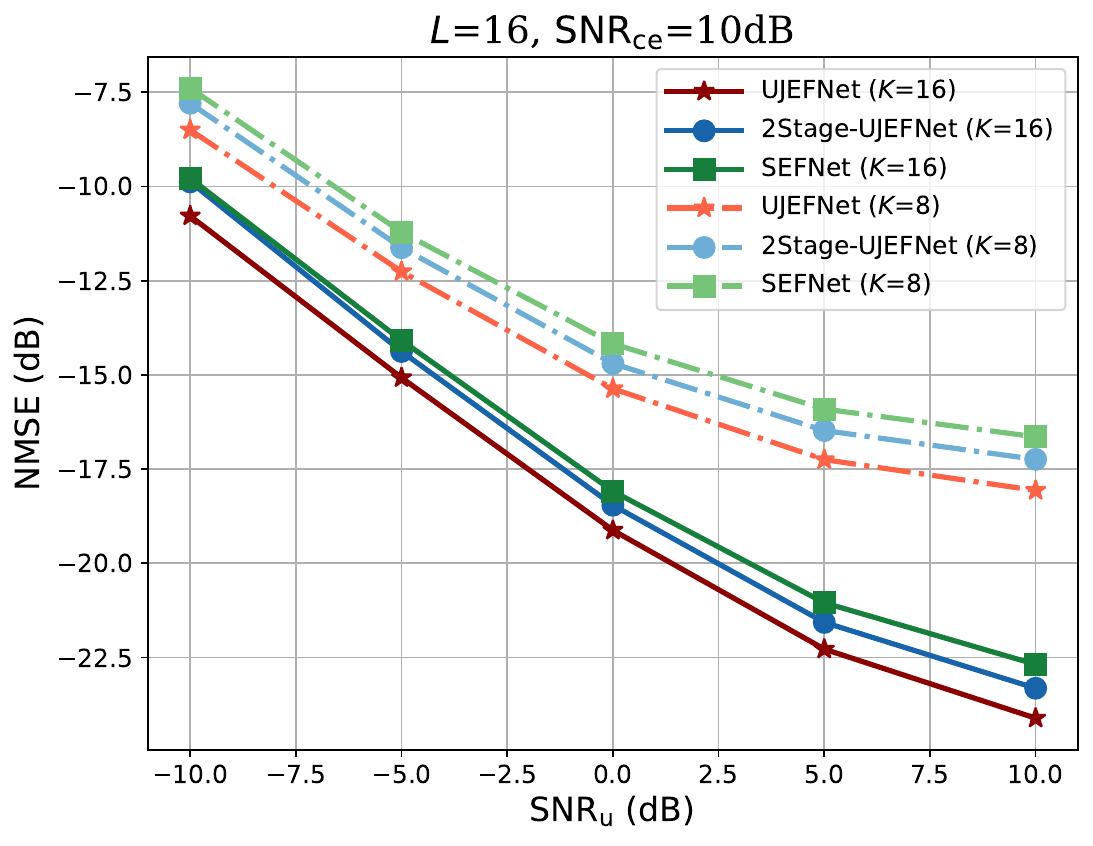}
	\caption{Ablation experiments: Evaluating the performance gains achieved through joint end-to-end training of multiple modules.}
	\label{Ideal_E2E_Advantages}
\end{figure}
To further validate the origin of the performance gain achieved by the proposed UJEFNet over the traditional downlink CE and CSI feedback architecture, sufficient ablation experiments are conducted. In UJEFNet, the ``\emph{Joint Refine}'' module leverages the partial correlation between uplink and downlink CSI to enhance CSI reconstruction accuracy compared to JEFNet. However, the introduction of the ``\emph{Joint Refine}'' module inevitably increases the depth of the decoder at the BS side. To determine whether the performance gain stems from the deeper network layers or the assistance of the reciprocity between the uplink and downlink CSI, we propose a network called DJEFNet for validation. DJEFNet adds a module to JEFNet with the same depth as the ``\emph{Joint Refine}'' module but removes the ``Concat'' layer, thereby eliminating the use of uplink CSI information for downlink CSI reconstruction. In this experiment, we set $L = 16$, $\mathrm{SNR_{ce}} = 10$ dB, and $K \in \{8, 16\}$. The experimental results are shown in Fig.~\ref{Ideal_Ablation}.

When $K = 8$, compared to JEFNet, UJEFNet achieves a performance gain of 1.10 dB at $\mathrm{SNR_u} = -10$ dB and 1.24 dB at $\mathrm{SNR_u} = 10$ dB. In contrast, DJEFNet only achieves a performance gain of 0.06 dB and 0.14 dB at $\mathrm{SNR_u} = -10$ dB and $\mathrm{SNR_u} = 10$ dB, respectively, despite the increased network depth. This indicates that the performance gain of UJEFNet primarily stems from the partial reciprocity between uplink and downlink CSI, which provides additional auxiliary information for improving downlink CSI reconstruction accuracy. Under the condition of $K = 8$, simply deepening the feedback network cannot effectively enhance the reconstruction performance.

When $K = 16$, UJEFNet achieves approximately 1.14 dB performance gain at both $\mathrm{SNR_u} = -10$ dB and $\mathrm{SNR_u} = 10$ dB compared to JEFNet. In contrast, DJEFNet achieves only 0.31 dB and 1.44 dB performance gain at $\mathrm{SNR_u} = -10$ dB and $\mathrm{SNR_u} = 10$ dB, respectively. This demonstrates that at low SNR, deepening the network has limited effect on improving channel reconstruction accuracy, while introducing uplink CSI assistance can bring stable performance gains. Although deepening the network at high SNR can achieve a performance gain close to that of uplink CSI-assisted reconstruction, this is due to the limited correlation between uplink and downlink CSI in FDD systems with a 200 MHz frequency gap and the simplicity of the ``\emph{Joint Refine}'' module, which is designed solely to validate the proposed thought.
\begin{figure*}[!t]
	\centering
	\includegraphics[width=6.8in]{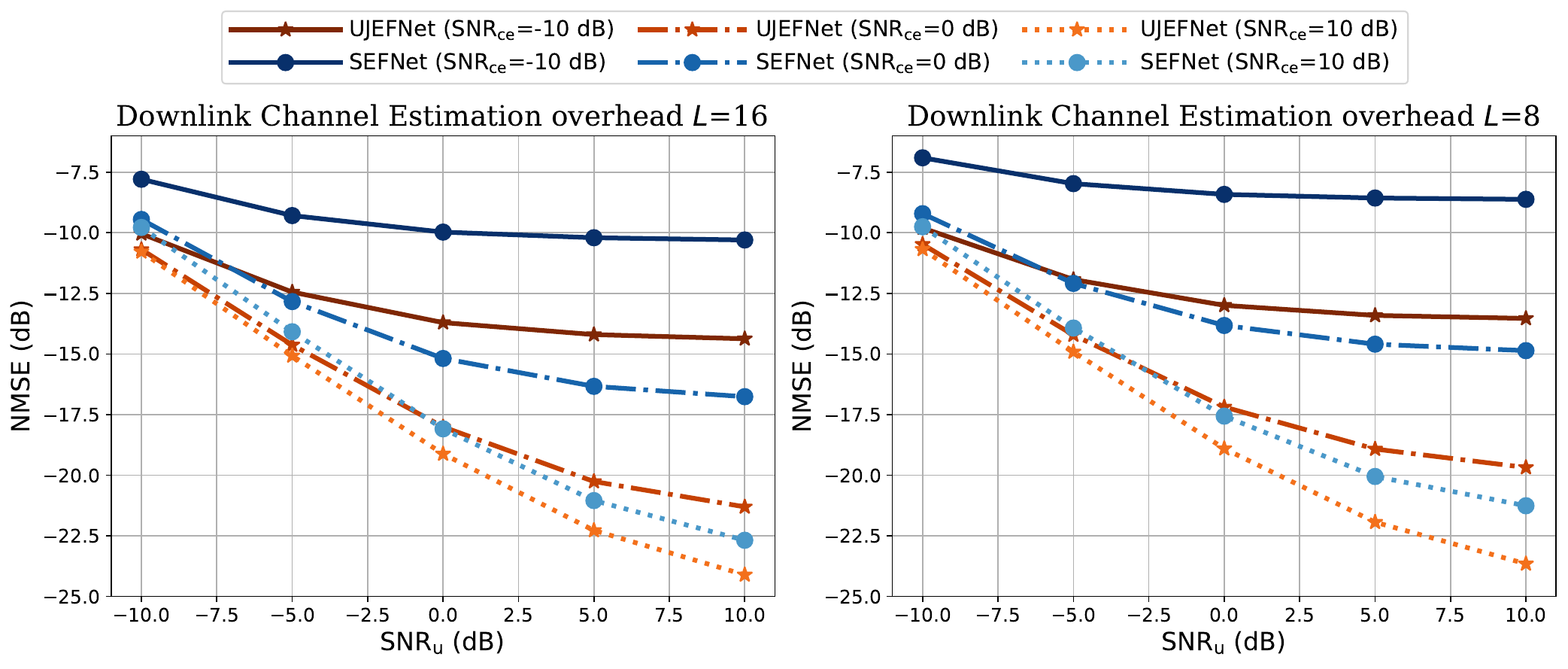}
	\caption{Scalability test: Evaluating the impact of CE accuracy on end-to-end performance during the downlink CE phase.}
	\label{Ideal_Scalability}
\end{figure*}

In this work, the entire network is trained in an end-to-end manner, including joint estimation and CSI compression, joint source-channel coding, and joint CSI reconstruction assisted by uplink CSI. The advantages and necessity of DJSCC have been demonstrated in Section \ref{EX_Ideal_CE}. Next, we aim to validate the necessity of joint CE and CSI compression, and CSI reconstruction with uplink assistance. Specifically, we replace the CSI feedback network under the SSCC architecture in Fig.~\subref*{SEFNet} with the CSI feedback network under the DJSCC architecture, i.e., the DJSCC architecture proposed in \cite{ADJSCC}. We first train the downlink CE module and the deep DJSCC-based CSI feedback module separately and then conduct joint testing, which we refer to as SEFNet in Fig.~\ref{Ideal_E2E_Advantages}. By comparing UJEFNet with SEFNet, we can assess the effectiveness of joint estimation with CSI compression. Furthermore, to verify the necessity of joint CSI reconstruction assisted by uplink CSI, we first train the JEFNet network and then concatenate it with the ``\emph{Joint Refine}'' module for end-to-end training. However, the network parameters of the JEFNet part are frozen to eliminate the influence of joint training. This architecture is referred to as ``2Stage-UJEFNet'' in Fig.~\ref{Ideal_E2E_Advantages}.

The experimental results in Fig.~\ref{Ideal_E2E_Advantages} demonstrate that joint estimation and feedback yield a performance gain of approximately 1 dB to 1.43 dB when $K = 16$ and $\mathrm{SNR_u} \in {-10 \text{ dB}, 10 \text{ dB}}$. For $K = 8$, the gain is around 1.1 dB to 1.42 dB. This performance improvement primarily stems from the fact that the source directly compressed by the UE encoder in UJEFNet is a lower-dimensional receiver pilot matrix, unlike SEFNet, which compresses the estimated full CSI. Additionally, a minor portion of the gain originates from the CSI compression module in UJEFNet, which is aware of the input bias caused by CE errors during the training phase, thereby mitigating the performance degradation caused by the mismatch between the training and testing distributions. However, since the pilot overhead of downlink CE is set to $L = 16$ and the downlink CE $\mathrm{SNR_{ce}} = 10$ dB, the downlink CE is relatively accurate at $-20.98$ dB, resulting in a minor distribution bias between the estimated CSI and the ideal CSI. The performance gap between ``UJEFNet'' and ``2Stage-UJEFNet'' reflects the gain achieved through joint CSI reconstruction assisted by uplink CSI. Specifically, for $K = 16$ and $\mathrm{SNR_u} \in {-10 \text{ dB}, 10 \text{ dB}}$, joint estimation and feedback can achieve a performance gain of 0.79 dB to 0.90 dB. For $K = 8$, the gain is around 0.70 dB to 0.82 dB. This gain is attributed to the end-to-end training strategy, which enables the transmitter to extract additional feature information that cannot be obtained from uplink CSI, thereby enhancing the retention of useful information.

\subsection{Scalability Analysis}

To investigate the impact of downlink CE pilot overhead and channel quality on the overall CSI reconstruction accuracy, we conducted a simulation experiment in this section. In the left subfigure of Fig.~\ref{Ideal_Scalability}, we set the pilot overhead for downlink CE to $L = 16$ and vary the SNR in the CE stage, i.e., $\mathrm{SNR_{ce}} \in \{-10 \text{ dB}, 0 \text{ dB}, 10 \text{ dB}\}$. The impact of the joint estimation and feedback architecture on the end-to-end performance is evaluated by comparing the CSI reconstruction accuracies of UJEFNet and SEFNet under different $\mathrm{SNR_u}$. In the right subfigure of Fig.~\ref{Ideal_Scalability}, we repeat the evaluation with $L = 8$.

From Fig.~\ref{Ideal_Scalability}, it can be observed that reducing the pilot overhead $L$ leads to performance degradation for both UJEFNet and SEFNet, with SEFNet being more severely impacted. Specifically, when $L$ is reduced from 16 to 8 under $\mathrm{SNR_{ce}} = -10$ dB and $\mathrm{SNR_u} \in \{-10 \text{ dB}, 10 \text{ dB}\}$, UJEFNet experiences a performance degradation of 0.24 dB to 0.84 dB, while SEFNet suffers a degradation of 0.88 dB to 1.68 dB. Under $\mathrm{SNR_{ce}} = 0$ dB, UJEFNet's performance drops by 0.21 dB to 1.62 dB, while SEFNet drops by 0.23 dB to 1.90 dB. When $\mathrm{SNR_{ce}} = 10$ dB, UJEFNet degrades by 0.10 dB to 0.45 dB, while SEFNet degrades by 0.03 dB to 1.42 dB.

When fixing $L = 16$, UJEFNet achieves a performance gain of 1.00 dB and 1.43 dB over SEFNet under $\mathrm{SNR_{ce}} = 10$ dB at $\mathrm{SNR_u} = -10$ dB and $\mathrm{SNR_u} = 10$ dB, respectively. When $\mathrm{SNR_{ce}} = 0$ dB, the performance gains increases to 1.24 dB and 4.54 dB, respectively. At $\mathrm{SNR_{ce}} = -10$ dB, UJEFNet achieves a gain of 2.25 dB and 4.07 dB, respectively. UJEFNet demonstrates superior performance at $L=8$, achieving a gain of 4.91 dB at $\mathrm{SNR_{ce}} = -10$ dB and $\mathrm{SNR_u} = 10$ dB compared to SEFNet. This is because, as the pilot overhead for CE decreases and the channel SNR becomes lower, the bias between the estimated downlink CSI and the ideal CSI becomes more significant, leading to a more substantial performance degradation in SEFNet. In contrast, UJEFNet is less sensitive to CE errors since this bias is already considered during the training phase.
\section{Conclusion}
\label{section4}
In this paper, we propose an uplink-assisted joint estimation and feedback network based on the DJSCC architecture, which is called UJEFNet. UJEFNet integrates multiple modules, including downlink pilot design, downlink CE, CSI compression, channel coding, channel decoding, and CSI reconstruction in traditional modular communication systems. Meanwhile, by leveraging the partial reciprocity between uplink and downlink channels in FDD systems, UJEFNet enhances CSI reconstruction accuracy, reduces feedback overhead, and improves spectrum efficiency.

% In the experimental section, we consider the impact of uplink CE errors, which are often overlooked in existing DJSCC architectures. We propose a training strategy that incorporates uplink CE errors during the training phase to enhance the accuracy of CSI reconstruction. Subsequently, we compare UJEFNet under the DJSCC architecture with different networks under the SSCC architecture to verify the necessity of DJSCC architecture, instead of SSCC architecture. The results show that UJEFNet under the DJSCC architecture can better capture the feature information of CSI for compression, achieving a performance gain of 9.2 dB to 14.0 dB within the tested SNR range. Furthermore, UJEFNet consistently outperforms SEFNet under the separated estimation and feedback architecture, especially in scenarios with low downlink CE accuracy. Additionally, we conduct extensive ablation experiments and scalability tests to analyze the sources of performance gains and validate the effectiveness of the proposed architecture.

In our experiments, we address uplink CE errors—often neglected in existing DJSCC frameworks—by proposing a training strategy integrating uplink CE errors to improve CSI reconstruction accuracy. Comparisons between DJSCC-based UJEFNet and SSCC-based networks validate DJSCC’s superiority, with UJEFNet achieving 9.2-14.0 dB gains within the tested SNR range by better capturing CSI features for compression. It also consistently surpasses SEFNet, particularly with low-accuracy downlink CE. Ablation studies and scalability tests further dissect performance sources and confirm the architecture’s effectiveness.

\bibliographystyle{IEEEtran}
\bibliography{myref}

% \newpage

 

% \end{IEEEbiography}



% \vfill

\end{document}